# Unifying the Gutenberg-Richter Law with Probabilistic Catalog Completeness

Jiawei Li[1*], Xinyi Wang[1,2], and Didier Sornette[1*]

1 Institute of Risk Analysis, Prediction and Management (Risks-X), Academy for Advanced Interdisciplinary Studies, Southern University of Science and Technology (SUSTech), Shenzhen, China.
2 School of Computer Science, Chengdu University of Information Technology (CUIT), Chengdu, China.

Corresponding authors: Didier Sornette (didier@sustech.edu.cn); Jiawei Li (lijw@cea-igp.ac.cn)

# Jiawei Li and Xinyi Wang contributed equally to this work

**Key Points:**

(1) We propose an augmented modeling framework that unifies the Gutenberg-Richter law with a probabilistic treatment of catalog incompleteness.
(2) We evaluate four variants of the proposed Gutenberg-Richter model using both synthetic and empirical catalogs across multiple performance metrics.
(3) The GR-AEReLU, allowing for an asymmetric crossover with two distinct exponential decays, consistently outperforms other variants and offers valuable seismological insights.
(4) The earthquake *b*-value varies systematically by region and tectonic setting, not the universal constant of 1.0 assumed in standard seismology.

**Abstract**
We propose a probabilistic approach to modeling catalog incompleteness through four candidate augmented Gutenberg-Richter (GR) laws, which incorporates incompleteness into the frequency-magnitude distribution (FMD) using two parameters, $m_c$, the transition magnitude, and $\sigma_c$, which defines the transition range from incompleteness to completeness. The four GR models are tested on synthetic and empirical catalogs, using multiple performance evaluation metrics. The GR-AEReLU model, which allows for an asymmetry in the convergence to the pure linear GR law for $m > m_c$ relative to the censorship of earthquakes of sizes smaller than $m_c$, is found to consistently outperform, providing more robust estimates of seismological parameters (e.g., $b$-value) that better reflect realistic physical conditions and observational characteristics. This augmented framework offers three main advantages: (1) unified modeling of incompleteness into the FMD, (2) parameters with clear physical and statistical meaning, and (3) the ability to capture nonlinear and asymmetric detection behaviors. Finally, our analysis reveals systematic regional variations in earthquake $b$-values that deviate significantly from the assumed universal value of 1.0, challenging a fundamental paradigm in seismology and demonstrating the need for region-specific values that reflect local tectonic conditions in seismic hazard assessments.

**Plain Language Summary**
Replacing the traditional sharp completeness threshold with a smooth probabilistic transition allows for a more realistic representation of earthquake detection probabilities, acknowledging both missed small events below the completeness magnitude and occasional omissions above it. In the present study, we propose a novel modeling approach that smoothly integrates the Gutenberg-Richter law with a probabilistic representation of catalog completeness. We evaluate four versions of this new Gutenberg-Richter model using both synthetic and real earthquake catalogs, across multiple performance metrics. Our results consistently show that the GR-AEReLU model, which introduces asymmetry crossover with two distinct exponential decays, provides the most accurate and reliable estimates of seismological parameters. These improved estimates align closely with realistic physical conditions and observed data characteristics. Our enhanced model offers significant advantages: (i) it naturally integrates magnitude distribution and completeness into a single unified framework; (ii) its parameters have clear physical meanings and intuitive statistical interpretations; and (iii) it accurately captures the nonlinear and asymmetric behaviors observed in real catalog. It can enhance seismicity models by incorporating small, previously discarded events with appropriate probabilistic weighting.

## 1. Introduction

The Gutenberg-Richter (GR) law, first proposed by Gutenberg & Richter (1944; 1954), is a fundamental empirical relationship in seismology that quantitatively characterizes the frequency-magnitude distribution (FMD) of earthquakes in a given region and within a given time window. Mathematically, it is expressed as $\log_{10}N(m) = a - bm$, where $N(m)$ represents the total number of earthquakes for a selected geographic area and during a specified period with magnitudes $\geq m$, and $a$ and $b$ are empirically derived parameters. Parameter $a$ measures the total number of observed earthquakes with $m \geq 0$. Parameter $b$ quantifies the relative proportion of large to small earthquakes. When normalized by the total number $N$ of earthquakes, it is nothing but an empirical sampling of the underlying complementary cumulative distribution function CCDF($m$), describing the probability to observe an earthquake with magnitudes $\geq m$. In other words, $N(m)/N$ converge to CCDF($m$) in the limit $N \rightarrow \infty$ according to the law of large numbers. The derivative of CCDF($m$) with respect to $m$ defines the probability density function PDF($m$), which is the likelihood that an earthquake has a specific magnitude $m$. Although the GR law describes earthquake magnitudes as following an exponential distribution, it translates into a power law distribution when expressed in terms of seismic moment or energy, reflecting the scale-invariant characteristics of seismic processes (Bak & Tang, 1989; Sornette & Sornette, 1989; Sornette et al., 1996). Due to its simplicity, robustness, and broad applicability, the GR law constitutes a fundamental cornerstone in seismic hazard assessment, earthquake forecasting, and the study of earthquake physics (Kagan et al., 1997; Geller et al., 1997; Klügel et al., 2006; Nandan et al., 2022; Mizrahi et al., 2024). Accurate characterization of the FMD is therefore critical not only for probabilistic seismic hazard analyses but also for gaining insights into the physical processes underlying earthquake occurrence and magnitude distribution.

In empirical earthquake catalogs, deviations from the expected linear shape of the GR law plotted on a logarithmic scale are commonly observed at lower magnitudes. Numerous studies have proposed adjusted GR models to characterize these deviations and have explored their physical underpinnings through the lenses of source physics and earthquake fractal characteristics. For instance, Aki (1987) observed that the departure of the empirical FMD from the GR law for earthquakes below magnitude 3 coincides with the widely observed deviation occurring at the moment-corner frequency from the $1/\omega^2$ spectrum of seismic waves expressing self-similarity. The corner frequency, characterizing the transition in the seismic wave spectrum, corresponds directly to a physical scale reflecting earthquake source properties. Several studies have introduced modified GR models to account for observed deviations in earthquake magnitude distributions, employing nonextensive statistical mechanics—a generalization of classical statistical mechanics suitable for complex systems with long-range interactions and fractal structures. For instance, Sotolongo-Costa & Posadas (2004) proposed the fragment-asperity interaction model, which incorporates the irregular geometry of fault planes and the fragments filling the gaps between them, suggesting that these fragments play a significant role in earthquake dynamics. An energy distribution function, which gives the GR law as a particular case, is analytically deduced by them based on the fragment size distribution function derived from a nonextensive formulation. Silva et al. (2006) revisited this model within the framework of Tsallis nonextensive statistics, proposing a new GR distribution function that differs from the original. Similarly, Darooneh & Mehri (2010) employed the $q$-exponential function, motivated by nonextensive statistical mechanics, to propose a modified GR relationship accommodating such deviations. More recently, da Silva (2021) introduced a $\kappa$-generalized GR law based on the $\kappa$-probability distribution, derived from Kaniadakis statistics.

However, interpretations linking such deviations directly to intrinsic earthquake dynamics may be premature, as these analyses often overlook the critical issue that departures from the GR law are more likely attributable to observational biases and incomplete earthquake catalogues rather than fundamental physical processes. With the advent of artificial intelligence (AI) techniques in seismic event identification and detection, the quality of earthquake catalogs has markedly improved (Kong et al., 2019; Mousavi & Beroza, 2022; 2023; Jia & Zhou, 2024; Kubo et al., 2024). Recent studies show that, in high-resolution catalogs, the magnitude at which deviations from the GR relationship appear has shifted to significantly lower values (e.g., Zhou et al., 2021; Ding et al., 2023; Li et al., 2024). This suggests that the observed deviations in the FMD are still primarily associated with missing earthquake records, underscoring the importance of considering catalog completeness in seismic analyses (Mignan & Woessner, 2012; Li et al., 2023; Wang et al., 2025).

The catalog completeness magnitude, denoted as $m_c$, is the magnitude above which all earthquakes in a region are assumed to be detected reliably (Rydelek & Sacks, 1989; Mignan & Woessner, 2012). However, this completeness threshold exhibits spatial and temporal variations influenced by factors such as changes in the seismic network, environmental noise, and regional seismicity characteristics (e.g., Mignan et al., 2013; Feng et al., 2022; Li et al., 2023). Most seismological models that utilize the GR law assume a fixed $m_c$, applying the GR law only above this threshold under the presumption of complete data for $m \geq m_c$. Typically, the catalog's incomplete portion ($m < m_c$) and complete portion ($m \geq m_c$; the GR law) are treated separately (e.g., Woessner & Wiemer, 2005; Mignan, 2012; García-Hernández et al., 2019; van der Elst, 2021). This binary threshold approach oversimplifies the complexities of real-world seismic monitoring, which is more accurately characterized by a gradual transition—from undetectability of very small earthquakes, to partial observability of intermediate events, and near-certain detection of larger earthquakes. This binary approach may lead to inaccuracies in seismic hazard assessments and statistical parameter estimations. Therefore, it is crucial to develop models that account for the probabilistic nature of earthquake detection and the dynamic variations in catalog completeness. Such models would provide a more accurate representation of the FMD across the entire spectrum of earthquake magnitudes, leading to improved seismic hazard assessments and a deeper understanding of earthquake processes.

Several probabilistic frameworks for estimating $m_c$ have been developed to overcome the limitations of fixed-threshold approaches, allowing for a more realistic representation of detection capabilities across varying magnitudes:

(1) Probabilistic Magnitude of Completeness (PMC) Method: Introduced by Schorlemmer & Woessner (2008), the PMC method estimates earthquake detection probabilities by analyzing empirical data, including phase picks and station metadata. While this approach offers a gradated understanding of detection capabilities, it requires extensive datasets beyond standard earthquake catalogs, such as detailed station information and attenuation relations, making its application complex and data intensive. Additionally, this method is challenging to integrate with other statistical seismological models, particularly the GR model, limiting its use primarily to assessments of catalog data quality and network monitoring capabilities.

(2) Ogata & Katsura (1993) Model (OK1993): This model employs a cumulative Normal distribution function (CDF), as recommended by Ringdal (1975), to represent the monotonically increasing probability of earthquake detection with magnitude and has clear physical interpretations related to instrument sensitivity and background noise. This function is then multiplied by the classical GR probability density function to construct a new magnitude distribution model (Ogata & Katsura, 1993; 2006). The model requires estimating parameters $\mu$ (the magnitude at which earthquakes are detected with 50% probability) and $\sigma$ (the size of magnitudes changes over which detection transitions from partial to complete), facilitating straightforward data fitting. Ogata & Katsura (1993) also observed that, when earthquakes below a certain threshold magnitude are excluded, a logistic sigmoid function provides a more realistic representation of the detection rate, with the completeness magnitude implicitly defined as $m_c = \mu + n\sigma$, where $n$ denotes the confidence level (e.g., $n = 0$ for 50% detection, $n = 1, 2, 3$ for approximately 68%, 95%, and 99% detection, respectively). However, the assumption of a Normal distribution imposes mirror symmetry in the detection probability curve, implying that the probability increases towards 1 for magnitudes much greater than $\mu$ at the same rate it decreases to 0 for magnitudes much smaller than $\mu$. This symmetry may not align with real-world observations, where detection probability often drops off steeply for smaller magnitudes but increases more gradually toward certainty for larger magnitudes. This imposed mirror symmetry could reduce the model's physical realism, as actual detection probabilities at lower magnitudes may exhibit more complex decay behaviors that a Normal CDF cannot accurately capture. Furthermore, the approach of Ogata & Katsura (1993) adopts a multiplicative or separable parameterization, effectively treating the determination of catalog completeness as independent from the statistical modeling of the GR distribution. This separation assumes absence of coupling between the functions and parameters governing detection and those defining earthquake occurrence rates, preventing a unified treatment of the two constructions that are likely to be coupled. While the assumed independence provides a natural and tractable starting point, it limits the model's ability to capture potential interdependencies in how catalogues imperfection impact the statistical determination of the frequency of earthquakes. Recognising the nature of the independence assumption in OK1993 makes it well-suited for improvement and generalization.

(3) Kijko & Smit (2017) Model: This model utilizes a power law function to represent the magnitude detection probability, which is then multiplied by the classical GR distribution to form a new distribution model. This approach accounts for the long tail characteristics of detection probability at lower magnitudes and features a flexible parameter structure, with parameters $\alpha$ and $\gamma$ capturing diverse detection probability behaviors, theoretically accommodating a broader range of empirical data patterns. However, the parameters $\alpha$ and $\gamma$ suffer from ambiguous interpretations and lack clear physical meanings—an issue also evident in modified GR relationships proposed by Sotolongo-Costa & Posadas (2004), Silva et al. (2006), Darooneh & Mehri (2010), and da Silva (2021). While power law functions offer significant flexibility, the selection of $\gamma$ may lack empirical justification. Additionally, the strong correlation between $\alpha$ and $\gamma$ (collinearity) can lead to instability and excessive sensitivity in parameter estimation. Similar to OK1993, this model employs a multiplicative approach that artificially separates completeness from the magnitude distribution, rather than capturing parameter interdependencies within a unified framework. As a result, it lacks a cohesive, concise, and intrinsically integrated structure that reflects the potentially coupled nature of the measure of earthquake frequency and detection processes.

In light of these limitations, we propose a novel enhanced GR model that offers the following advantages: (i) a unified framework that naturally combines catalog completeness and magnitude distribution within a single expression; (ii) parameters with clear physical definitions and intuitive statistical interpretations; and (iii) the capability to capture the nonlinear and asymmetric characteristics of magnitude detection probability, accurately reflecting real-world scenarios such as rapid decay in detection probability and asymmetry at lower magnitudes observed in earthquake catalog data.

## 2. Four augmented forms of the Gutenberg-Richter law using gReLU functions

Earthquake catalogs often include incomplete portions, which complicates and may bias their analysis. One key challenge is accurately representing the complementary cumulative frequency-magnitude distribution (CCFMD) of earthquakes recorded in these catalogs. To address this, we propose an augmented GR law tailored to handle incompleteness, expressed as:

$$\mathrm{CCFMD}(m) = 10^{a - b \cdot G(m)} \quad (1)$$

in which,

$$G(m) = m_\mathrm{c} + \sigma_\mathrm{c} \cdot \mathrm{gReLU}(x) \quad (2)$$

where $x = (m - m_\mathrm{c})/\sigma_\mathrm{c}$. This enhanced model augments the traditional GR law by integrating the determination of the FMD with a detailed description of catalog completeness. Unlike standard approaches, which treat completeness as a separate issue or an add-on analysis, this formulation seamlessly incorporates completeness into the model itself. Specifically, it extends the classical parameters of the GR law—the $a$-value and $b$-value—by introducing two additional parameters: $m_\mathrm{c}$ replaces the completeness magnitude as a transition magnitude around which earthquake detection change rapidly, and $\sigma_\mathrm{c}$ quantifies the magnitude range over which this transition occurs.

The generalized function gReLU($x$) is required to approach gReLU($x$) = 0 as $x \to -\infty$ and to converge to gReLU($x$) = $x$ as $x \to +\infty$ in order to recover the standard GR distribution for large magnitudes. We do not consider here the issue of tapering or potential outliers in the GR distribution at very large magnitudes, as discussed in Sornette & Sornette (1999), Kagan (2002a; 2002b), and Li et al. (2025a; 2025b). A specific instance of this function is the Rectified Linear Unit, ReLU($x$), a widely used activation function in deep learning that introduces nonlinearity into models (Dayan & Abott, 2001; Nair & Hinton, 2010). The ReLU($x$) function is defined as ReLU($x$) = 0 for $x < 0$ and ReLU($x$) = $x$ for $x \geq 0$ (Figure 1a). When gReLU($x$) is set equal to ReLU($x$), the resulting CCFMD($m$) becomes $10^{a-bm}$ for $m \geq m_c$ and the constant $10^{a-bm_c}$ for $m < m_c$ (Figure 1b). This corresponds to the exact GR law for $m \geq m_c$, with no earthquakes considered for $m < m_c$, as CCFMD($m$) remains constant in this range. Here, we aim to seek functions that provide a smooth transition between the GR law for $m \geq m_c$ and a gradual decrease in recorded earthquake frequencies for $m < m_c$. To satisfy these requirements, the smoothing function must not only converge to 0 as $x \to -\infty$ and to $x$ as $x \to +\infty$, but it must also ensure that CCFMD($m$) remains monotonic. Consequently, gReLU($x$) must be a monotonically increasing function of $x$. Below are four gReLU($x$) functions that meet these criteria:

(1) **SSReLU**. A function that satisfies the above requirements is the standard Softplus Function, which is commonly used in machine learning and neural networks (Liu & Furber, 2016; Wiemann et al., 2021). In this study, we refer to it as SSReLU, and it is defined by:

$$\mathrm{SSReLU}(x) = \ln(1 + e^x) \tag{3}$$

(2) **BSReLU**. Another function that meets all these requirements is inspired by the formula for calculating the price of a call option according to the Black-Scholes model (Black & Scholes, 1973; Bouchaud and Sornette, 1994) in financial economics. In this study, we refer to it as BSReLU, which is defined as follows:

$$\begin{aligned}\mathrm{BSReLU}(x) = &\left(x + \frac{m_c}{\sigma_c}\right)\Phi\left(\frac{\ln\left[\left(x\sigma_c + m_c\right)/m_c\right]}{\sigma_c} + \frac{\sigma_c}{2}\right)\\ &-\frac{m_c}{\sigma_c}\Phi\left(\frac{\ln\left[\left(x\sigma_c + m_c\right)/m_c\right]}{\sigma_c} - \frac{\sigma_c}{2}\right)\end{aligned} \tag{4}$$

In terms of the variable $x = (m - m_c)/\sigma_c$, this corresponds to

$$\begin{aligned}\mathrm{BSReLU}\left(\frac{m - m_c}{\sigma_c}\right) = &\frac{m}{\sigma_c}\Phi\left(\frac{\ln\left(m/m_c\right)}{\sigma_c} + \frac{\sigma_c}{2}\right)\\ &-\frac{m_c}{\sigma_c}\Phi\left(\frac{\ln\left(m/m_c\right)}{\sigma_c} - \frac{\sigma_c}{2}\right)\end{aligned}$$

Here, $\Phi$ denotes the standard Normal cumulative distribution function. In the Supporting Information of our separate study (Wang et al., 2025), we show that two independent stochastic constructions, a lognormal exceedance model and a geometric Brownian motion, lead to the same closed-form expression for the BSReLU function.

(3) **COReLU**. Unlike the previous two continuous functions, we also introduce a piecewise function with a cut-off property, which we refer to as COReLU, defined as follows:

$$\mathrm{COReLU}(x) = \begin{cases} e^{x-1} & x < 1 \\ x & x \geq 1 \end{cases} \tag{5}$$

This function is composed of an exponential function and a linear function. Unlike ReLU($x$), which consists of two distinct linear segments joined at $x$ = 0, COReLU($x$) consists of one nonlinear curve and one linear segment joined at $x$ = 1. This design ensures a smooth transition of the function while avoiding a step discontinuity at the cut-off point. In other words, COReLU($x$) and its first derivative are continuous at $x$ = 1.

(4) **AEReLU**. This function allows for an asymmetric exponential convergence to 0 for $x < 0$ and to $x$ for $x > 0$ where $\beta$ controls the strength of the asymmetry:

$$\mathrm{AEReLU}(x) = \begin{cases} \frac{e^{x}}{1+\beta} & x < 0 \\ x + \frac{e^{-\beta x}}{1+\beta} & x \geq 0 \end{cases} \tag{6}$$

This parameterization again ensures continuity and differentiability at the transition point $x$ = 0.

In terms of the variable $x = (m - m_c)/\sigma_c$, in the limit $\sigma_c \rightarrow 0$, SSReLU[$(m - m_c)/\sigma_c$], BSReLU[$(m - m_c)/\sigma_c$], COReLU[$(m - m_c)/\sigma_c$], and AEReLU[$(m - m_c)/\sigma_c$] converge to ReLU($m - m_c$), thus recovering the pure GR for $m \geq m_c$ and a constant for $m < m_c$. In summary, the SSReLU function models a symmetric crossover controlled by exponential decay, whereas the BSReLU function employs a symmetric crossover governed by a Gaussian profile. The COReLU function features exponential tapering only on the side of incompleteness. The AEReLU function extends SSReLU by introducing asymmetry through an additional parameter $\beta$, allowing for an asymmetric crossover controlled by two possibly different exponential decays.

By substituting the smoothing function gReLU($x$) from Equations (3), (4), (5), and (6) into Equations (1) and (2), we obtain four distinct augmented GR models. These models are referred to as GR-SSReLU, GR-BSReLU, GR-COReLU, and GR-ASReLU, respectively. It can be observed that GR-SSReLU, GR-BSReLU, and GR-COReLU share the same degrees of freedom, meaning they are characterized by an identical set of fitting parameters, $\theta = \{a, b, m_c, \sigma_c\}$. GR-AEReLU introduces an additional parameter, expanding its parameter set to $\theta = \{a, b, m_c, \sigma_c, \beta\}$. It is also worth noting that, by definition, the BSReLU function is only valid when both $m$ and $m_c$ are positive. Therefore, if the catalog to be fitted contains events with magnitudes less than zero, a shift parameter $C$ should be added to both $m$ and $m_c$ in Equations (2) and (4). In such cases, the GR-BSReLU model involves one additional fitting parameter, namely $C$. The fitting procedure is detailed in Supporting Information Text

S1. Figure 1 illustrates the four gReLU($x$) functions and their corresponding CCFMD($m$) representations when applied to Equations (1) and (2).

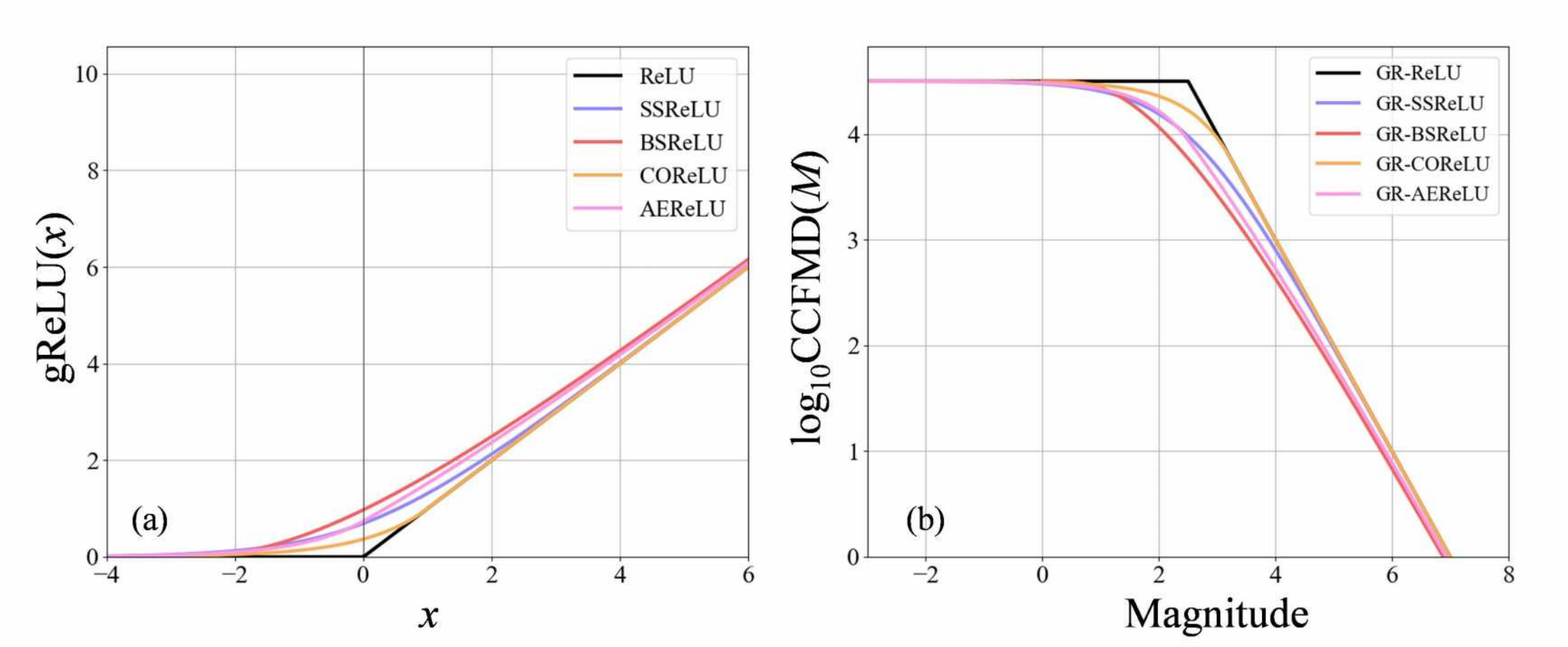

**Figure 1.** Plots of (a) the four gReLU($x$) functions introduced in the text and (b) the corresponding augmented complementary cumulative frequency-magnitude distributions (CCFMDs). For comparison, the figure also includes the Rectified Linear Unit, ReLU($x$), depicted with a black solid line, which is defined as ReLU($x$) = 0 for $x < 0$ and ReLU($x$) = $x$ for $x \geq 0$, along with the corresponding CCFMD recovering the pure Gutenberg-Richter law. The relationship between $x$ in (a) and the magnitude $m$ in (b) is given by $x = (m - m_c)/\sigma_c$. The parameters are set to $a = 7.0$, $b = 1.0$, $m_c = 2.5$, $\sigma_c = 0.75$, and $\beta = 0.35$, which are values that are informed by empirical data.

## 3. Earthquake catalog used for testing

We evaluate the augmented GR model, incorporating the gReLU($x$) functions, using both simulated earthquake catalogs and empirical earthquake catalogs. For the simulated earthquake catalogs, events are generated based on PDFs derived from each of the augmented GR models incorporating the four candidate gReLU($x$) functions. The PDFs have a minimum magnitude of $m_{\min} = 0$ and a maximum magnitude of $m_{\max} = 8.0$. The parameters used are $b^{\text{true}} = 1.0$, $m_c^{\text{true}} = 2.5$, $\sigma_c^{\text{true}} = 0.75$, and $\beta^{\text{true}} = 0.35$, with the total number of earthquakes set to $10^k$ for $k = 3, 4, 5, 6$. The true $a$-value $a^{\text{true}}$ is calculated as $a^{\text{true}} = k + 2.5 + 0.75\text{gReLU}(-10/3)$, based on Equations (1) and (2). Specifically, SSReLU(−10/3) ≈ 0.03505, BSReLU(−10/3) ≈ 0, COReLU(−10/3) ≈ 0.01312, and AEReLU(−10/3) ≈ 0.02378. The selection of these parameters is based on typical values observed in real earthquake catalogs (e.g., Table 1). These setups result in a total of 16 synthetic datasets, as each model generates data for four different values of $k$. For each combination of model and $k$, 200 independent datasets are sampled directly from the specified PDF, yielding 200 parallel realizations per group for subsequent performance evaluation and analysis.

The observed earthquake catalogs are selected from four regions in China and two additional regions, California and New Zealand, totaling six study regions. In China, the regions include Beijing-Tianjin-Hebei, the southeastern coast, Sichuan-Yunnan, and northern Xinjiang. Their boundaries are defined by the $m_c \approx 1.8$ contour lines from the posterior $m_c$ map generated using the Bayesian Magnitude of Completeness (BMC) method in Li et al. (2023). Data were sourced from the China Earthquake Networks Center (CENC) for the period January 1, 2009, to November 18, 2024. The California catalog, from the ANSS Comprehensive Earthquake Catalog (ComCat; U.S. Geological Survey, 2017), spans January 1, 2009, to November 2, 2023. Its boundary follows the $m_c \approx 1.8$ contour lines derived from the BMC method in Tormann et al. (2014). The New Zealand catalog, from the GeoNet Earthquake Catalog (GNS Science, 2022), covers January 1, 2009, to December 12, 2023. Its boundary is based on Petersen (2011), focusing on areas with dense and evenly distributed seismic stations. Figure S1 presents the magnitude-time sequence plots of these empirical earthquake catalogs from the six study regions.

## 4. Results of model evaluation

Tables S1(a) to S4(a) show the fitted parameters and performance evaluation metrics of the augmented Gutenberg-Richter model incorporating the four gReLU($x$) functions, applied to simulated catalogs generated by the GR-SSReLU, GR-BSReLU, GR-COReLU, and GR-AEReLU models, respectively, using predefined parameters (see Section 3 for details). Tables 1 and S5(a) show the fitted parameters and performance evaluation metrics of the augmented Gutenberg-Richter models applied to empirical earthquake catalogs from six study regions. To systematically assess model performance, we employ four evaluation metrics—root mean square error (Rmse), coefficient of determination (R-Square), sum of squared errors (SSE), and Akaike Information Criterion (AIC; Akaike, 2011)—to determine which of the four augmented Gutenberg-Richter models performs best under each scenario. Table S5(a) also report the $p$-values for empirical earthquake catalogs. Here, the $p$-value represents the probability that synthetic catalogs generated by the model show a poorer fit than the empirical catalog. Consequently, larger $p$-values indicate that the model provides an adequate fit to the empirical data, while smaller $p$-values suggest a poorer fit, potentially warranting model rejection. Tables S1(b) to S5(b) indicate the rankings of GR-PSReLU, GR-BSReLU, GR-COReLU, and GR-AEReLU based on these metrics, as presented in Tables S1(a) to S5(a), where a rank of "First" denotes the best-performing model for a given metric.

Figure 2 shows the proportion of times each model ranked as best ($1^{st}$, dark red), good ($2^{nd}$, light red), poor ($3^{rd}$, light blue), or worst ($4^{th}$, dark blue) for simulated and empirical catalogs. In Figure 2, Panels (II-i) and (II-ii) present alternative views of the same data from Table S5: (II-i) shows, for each metric, how often each model ranks 1st to 4th across the six study regions, while (II-ii) shows, for each region, how often each model ranks 1st to 4th across the five evaluation metrics. Among the four candidate models, GR-AEReLU demonstrates the greatest robustness and is therefore recommended as the preferred choice. For detailed analyses and comparisons of all models, we refer to Supporting Information Texts S2 to S4. In the cases of correct model specification — cases where each model is fitted to data generated by itself, GR-AEReLU exhibits favorable performances (1st and 2nd place). In the cases of model mis-specification — cases where a model is fitted to data generated by a different model, GR-AEReLU also consistently demonstrates greater robustness, ranking as best or good in at least 70% of cases, with no occurrences of being ranked as worst. These results indicate the superior adaptability of GR-AEReLU.

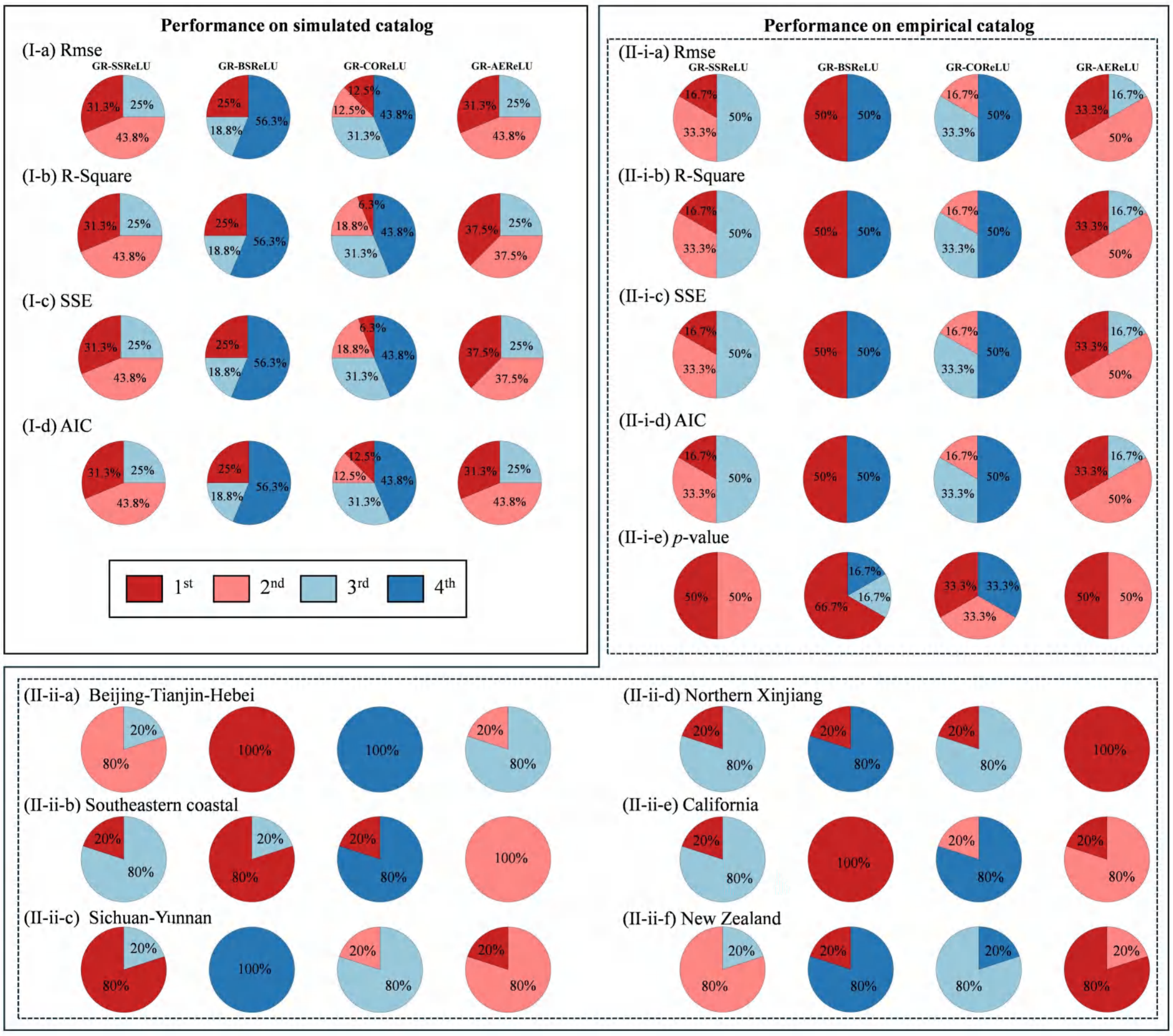

Figure 2. Proportions of times each model (GR-SSReLU, GR-BSReLU, GR-COReLU, and GR-AEReLU) ranked as best (1st, dark red), good (2nd, light red), poor (3rd, light blue), or worst (4th, dark blue) across different evaluation settings. Panels (I) summarize model rankings based on four metrics (Rmse, R-Square, SSE, and AIC) for simulated catalogs and five metrics (including *p*-value) for empirical catalogs. Panels (II-i) and (II-ii) present alternative views of the same data from Table S5: (II-i) shows, for each metric, how often each model ranks 1st to 4th across the six study regions, while (II-ii) shows, for each region, how often each model ranks 1st to 4th across the five evaluation metrics. Detailed rankings and metric values for each model are provided in Tables S1 to S5.

In Figure 2, GR-AEReLU stands out by attaining best or good rankings significantly more often than the other three models for the empirical catalogs from the six study regions. The GR-AEReLU model demonstrates consistently strong performance across diverse regions (except Beijing-Tianjin-Hebei), affirming the model's flexibility and robustness in capturing realistic earthquake completeness and magnitude distributions. Figures S2 to S6 display the probability density functions (PDFs) of the estimated values of $m_c^{pred}$, $\sigma_c^{pred}$, $a^{pred}$, $b^{pred}$, and $\beta^{pred}$, respectively, obtained from 200 bootstrap iterations. Figures S7 to S10 show CCFMD$^{obs}$($m$) of the empirical earthquake catalogs alongside with the fitted CCFMD$^{fit}$($m$) derived using the augmented GR mdoels. Figure S11 shows residuals between CCFMD$^{obs}$($m$) and CCFMD$^{fit}$($m$), expressed as $\log_{10}$CCFMD$^{fit}$($m$) - $\log_{10}$CCFMD$^{obs}$($m$).

**Table 1.** Fitted parameters of the augmented Gutenberg-Richter model with four gReLU($x$) functions applied to empirical earthquake catalogs from six study regions.

| Region | Model | Fitted parameters | | | | |
|---|---|---|---|---|---|---|
| | | $a^{pred}$ | $b^{pred}$ | $m_c^{pred}$ | $\sigma_{mc}^{pred}$ | $\beta^{pred}$ |
| Beijing-Tianjin-Hebei | **GR-SSReLU** | 5.48 ± 0.07 | 0.95 ± 0.03 | 0.69 ± 0.05 | 0.32 ± 0.03 | -- |
| | **GR-BSReLU** | 5.85 ± 0.42 | 1.03 ± 0.07 | 0.62 ± 0.43 | 0.73 ± 0.21 | -- |
| | **GR-COReLU** | 5.42 ± 0.04 | 0.93 ± 0.02 | 0.64 ± 0.03 | 0.50 ± 0.03 | -- |
| | **GR-AEReLU** | 5.56 ± 0.12 | 0.97 ± 0.04 | 0.76 ± 0.09 | 0.38 ± 0.04 | 0.56 ± 0.11 |
| Southeastern coastal | **GR-SSReLU** | 4.97 ± 0.08 | 0.93 ± 0.04 | 0.51 ± 0.06 | 0.28 ± 0.04 | -- |
| | **GR-BSReLU** | 5.62 ± 0.78 | 1.05 ± 0.11 | 0.74 ± 0.62 | 0.93 ± 0.28 | -- |
| | **GR-COReLU** | 4.93 ± 0.06 | 0.92 ± 0.03 | 0.46 ± 0.05 | 0.42 ± 0.05 | -- |
| | **GR-AEReLU** | 5.06 ± 0.13 | 0.96 ± 0.05 | 0.58 ± 0.10 | 0.32 ± 0.06 | 0.51 ± 0.10 |
| Sichuan-Yunnan | **GR-SSReLU** | 6.48 ± 0.03 | 0.86 ± 0.01 | 0.81 ± 0.02 | 0.21 ± 0.01 | -- |
| | **GR-BSReLU** | 6.79 ± 0.02 | 0.92 ± 0.02 | 1.10 ± 0.01 | 0.95 ± 0.07 | -- |
| | **GR-COReLU** | 6.48 ± 0.02 | 0.86 ± 0.01 | 0.81 ± 0.01 | 0.41 ± 0.01 | -- |
| | **GR-AEReLU** | 6.72 ± 0.14 | 0.87 ± 0.01 | 1.09 ± 0.16 | 0.39 ± 0.02 | 0.39 ± 0.72 |
| Northern Xinjiang | **GR-SSReLU** | 5.79 ± 0.07 | 0.89 ± 0.03 | 1.13 ± 0.05 | 0.21 ± 0.03 | -- |
| | **GR-BSReLU** | 6.08 ± 0.09 | 0.95 ± 0.03 | 1.36 ± 0.07 | 0.62 ± 0.09 | -- |
| | **GR-COReLU** | 5.78 ± 0.06 | 0.89 ± 0.02 | 1.12 ± 0.04 | 0.41 ± 0.03 | -- |
| | **GR-AEReLU** | 6.14 ± 0.14 | 0.92 ± 0.06 | 1.47 ± 0.08 | 0.39 ± 0.03 | 0.11 ± 0.20 |
| California | **GR-SSReLU** | 6.39 ± 0.09 | 1.13 ± 0.03 | 0.84 ± 0.05 | 0.35 ± 0.04 | -- |
| | **GR-BSReLU** | 6.94 ± 0.38 | 1.23 ± 0.08 | 1.10 ± 0.31 | 0.88 ± 0.17 | -- |
| | **GR-COReLU** | 6.27 ± 0.05 | 1.09 ± 0.02 | 0.76 ± 0.03 | 0.49 ± 0.04 | -- |
| | **GR-AEReLU** | 6.55 ± 0.14 | 1.17 ± 0.04 | 0.96 ± 0.09 | 0.41 ± 0.05 | 0.46 ± 0.07 |
| New Zealand | **GR-SSReLU** | 7.16 ± 0.06 | 1.03 ± 0.02 | 1.78 ± 0.03 | 0.24 ± 0.02 | -- |
| | **GR-BSReLU** | 7.45 ± 0.09 | 1.09 ± 0.02 | 1.95 ± 0.05 | 0.37 ± 0.04 | -- |
| | **GR-COReLU** | 7.13 ± 0.04 | 1.03 ± 0.01 | 1.76 ± 0.02 | 0.42 ± 0.01 | -- |
| | **GR-AEReLU** | 7.63 ± 0.10 | 1.11 ± 0.02 | 2.07 ± 0.06 | 0.39 ± 0.01 | 0.24 ± 0.16 |

## 5. Discussion

The superior performance demonstrated by GR-AEReLU model, which incorporates the additional parameter $\beta$ to account for the asymmetric crossover between the pure linear GR law for $m \geq m_c^{pred}$ and the incomplete part of the catalogues of earthquakes of sizes smaller than $m_c^{pred}$, underscores the critical importance of explicitly accounting for asymmetry in completeness modeling. Unlike symmetric functions such as GR-SSReLU and GR-BSReLU, or the single-sided exponential tapering exhibited by GR-COReLU, the GR-AEReLU model's asymmetry provides enhanced flexibility. This flexibility enables a more accurate and realistic representation of inherently asymmetric detection probabilities across various magnitude ranges, resulting in improved completeness modeling, which should translate in more reliable seismic hazard assessments. Our probabilistic approach offers a significantly more realistic and comprehensive characterization of earthquake catalogs, emphasizing the necessity of integrating region-specific detection attributes into statistical seismic modeling frameworks.

In this section, we primarily use the best-performing GR-AEReLU model to illustrate the concept of magnitude-dependent catalog completeness probability as defined by our proposed framework, as well as the asymmetric completeness characteristics revealed by the varying $\beta^{pred}$-values across regions. In addition, we introduce a new method that assigns different weights to the $b^{pred}$-values estimates from the four augmented GR models based on their AIC scores, enabling a unified, weighted $b^{pred}$-values estimate. This integrated estimate is then used to analyze and compare the seismicity characteristics of the six study regions.

## 5.1 Magnitude-dependent catalog completeness probability

In practice, the catalog completeness probability $P(m)$—that is, the probability that events of magnitude $m$ or larger are recorded in a given region—can be defined as the ratio between Equation (1) and the linear Gutenberg-Richter law, expressed as:

$$P(m) = \frac{10^{a-b\cdot G(m)}}{10^{a-b\cdot m}} = 10^{b[m-G(m)]} \tag{7}$$

Substituting Equation (2), along with $x = (m - m_c)/\sigma_c$, into Equation (7), we obtain:

$$P(m) = 10^{b\sigma_c\left[\frac{m-m_c}{\sigma_c} - \mathrm{gReLU}\left(\frac{m-m_c}{\sigma_c}\right)\right]} \tag{8}$$

Thus, by substituting the gReLU($x$) expressions from Equations (3) to (6) and the corresponding parameter estimates (Table 1) into Equation (8), one can compute the completeness probability for any given magnitude $m$.

Based on the GR-AEReLU estimates, we computed completeness probabilities at $m = m_c^{pred} + \sigma_c^{pred}$, $m = m_c^{pred} + 2\sigma_c^{pred}$, and $m = m_c^{pred} + 3\sigma_c^{pred}$ for all six study regions. In Beijing-Tianjin-Hebei, the completeness probabilities at $m = m_c^{pred} + \sigma_c^{pred} = 1.14$, $m = m_c^{pred} + 2\sigma_c^{pred} = 1.52$, and $m = m_c^{pred} + 3\sigma_c^{pred} = 1.90$ are 73.3%, 83.7%, and 90.4%, respectively. The southeastern coastal region performs comparably, with probabilities ranging from 75.5% at $m = m_c^{pred} + \sigma_c^{pred} = 0.90$ to 90.4% at $m = m_c^{pred} + 3\sigma_c^{pred} = 1.54$. These results indicate that these two regions possess the highest earthquake detection quality. Sichuan-Yunnan and California perform slightly worse. In Sichuan-Yunnan, the completeness probabilities increases from 68.3% at $m = m_c^{pred} + \sigma_c^{pred} = 1.48$ to 84.0% at $m = m_c^{pred} + 3\sigma_c^{pred} = 2.26$, while in California, probabilities range from 62.0% at $m_c^{pred} + \sigma_c^{pred} = 1.37$ to 82.7% at $m_c^{pred} + 3\sigma_c^{pred} = 2.19$. The poorest observational completeness is found in northern Xinjiang and New Zealand. In northern Xinjiang, the completeness probabilities are 51.3%, 55.0%, and 58.6% at $m_c^{pred} + \sigma_c^{pred} = 1.86$, $m_c^{pred} + 2\sigma_c^{pred} = 2.25$, and $m_c^{pred} + 3\sigma_c^{pred} = 2.64$, respectively. In New Zealand, the corresponding values are 53.3%, 60.8%, and 67.6% at $m_c^{pred} + \sigma_c^{pred} = 2.46$, $m_c^{pred} + 2\sigma_c^{pred} = 2.85$, and $m_c^{pred} + 3\sigma_c^{pred} = 3.24$. The limited performance in northern Xinjiang is primarily due to the sparse station coverage (e.g., Li et al., 2023), while in New Zealand it is attributed to the high proportion of earthquakes occurring at depths exceeding 30 km (Petersen, 2011).

## 5.2 Asymmetry in the transition from incompleteness to completeness

The estimated $\beta^{pred}$ values for GR-AEReLU are smaller than 1 across all six regions (Table 1), reflecting both regional variability in detection completeness and asymmetry in the transition around $m_c^{pred}$. Specifically, $\beta^{pred} < 1$ implies a slower convergence to the linear Gutenberg-Richter law for $m \geq m_c^{pred}$, relative to the suppression of earthquakes with magnitudes below $m_c^{pred}$. Beijing-Tianjin-Hebei (0.56 ± 0.11), southeastern coastal (0.51 ± 0.10), and California (0.46 ± 0.07) exhibit relatively larger $\beta^{pred}$-values with small uncertainties. This slower transition in the $m \geq m_c^{pred}$ region where earthquakes are more numerous may explain the smaller uncertainties in the parameter estimations. Sichuan-Yunnan (0.39 ± 0.72) exhibits a moderate $\beta^{pred}$-value with largest uncertainties. The rather large estimation uncertainty is influenced by statistics, as there is less data to constrain the faster transition in the domain $m \geq m_c^{pred}$. New Zealand (0.24 ± 0.16) and northern Xinjiang (0.11 ± 0.20) obtain the lowest $\beta^{pred}$, indicating a very gradual completeness transition. At magnitude $m = m_c^{pred} + 3\sigma_c^{pred} = 2.64$ and 3.24 for northern Xinjiang and New Zealand, respectively, approximately 41.4% and 32.4% of earthquakes are still estimated to be missing.

These results demonstrate the importance of this class of probabilistic models in which $m_c$ is no longer considered a strict cutoff that abruptly separates the complete from the incomplete regions of the CCFMD. Our result challenges the common assumption that a fixed completeness magnitude is adequate, showing instead that catalog incompleteness is more gradual since larger earthquakes can still be missing. Overall, our results indicate that GR-AEReLU provides the more accurate, informative and reliable estimates of seismological parameters among the models considered.

## 5.3 Seismicity characterization across regions: Insights from a multi-model augmented GR framework

The differences in the $b^{pred}$-values estimated by the four augmented GR models proposed in the present study are often larger than their respective standard deviations (Table 1; Figure S3). This indicates that the differences are not negligible and may carry meaningful information. In some cases—such as GR-SSReLU, GR-COReLU, and GR-AEReLU in the Sichuan–Yunnan region—the estimated $b^{\mathrm{pred}}$-values are identical or nearly identical, suggesting that the consistent estimates across models reflect a precise and reliable value of $b$-value. To take advantage of this variability while accounting for the differing levels of model reliability (i.e., explanatory power for the observed data), we compute a weighted average of the $b^{\mathrm{pred}}$-value PDFs from the four augmented GR models. This yields a unified, weighted estimate of the regional $b$-value, expressed as:

$$\ln[\mathrm{PDF}_{\mathrm{weight}}(b^{\mathrm{pred}})] = \frac{1}{4}\sum_{i=1}^{4}\omega_i \ln[\mathrm{PDF}_i(b^{\mathrm{pred}})] \tag{9}$$

where $\omega_i$ is the weight assigned to model $i$. To calculate these weights, we adopt the AIC in Table S5(a), which considers both goodness-of-fit and model complexity by penalizing overparameterization. The AIC-based weight is given by:

$$\omega_i = \frac{\exp(-\frac{1}{2}\Delta\mathrm{AIC}_i)}{\sum_{j=1}^{4}\exp(-\frac{1}{2}\Delta\mathrm{AIC}_j)} \tag{10}$$

with $\Delta\mathrm{AIC}_i = [\mathrm{AIC}_i - \min(\mathrm{AIC})]/[\max(\mathrm{AIC}) - \min(\mathrm{AIC})]$. This approach is analogous to Bayesian inference, where a posterior distribution is obtained by combining a prior and a likelihood. Here, the weighted average of multiple PDFs is obtained using a geometric weighting scheme derived from AIC differences. In principle, this procedure can also be applied to other parameter estimates beyond $b^{\mathrm{pred}}$-values. The resulting combined distributions represent our most robust and comprehensive parameter estimates.

Figure 3 shows the weighted averages of the $b^{\mathrm{pred}}$-value PDFs from the four augmented GR models for the six study regions (see also Figure S3 and Table 1). Among the six regions, Sichuan-Yunnan and northern Xinjiang yield mean $b^{\mathrm{pred}}$-values below 1.0, while mean $b^{\mathrm{pred}}$-values for the southeastern coast and Beijing-Tianjin-Hebei are closest to 1. In contrast, California and New Zealand exhibit mean $b^{\mathrm{pred}}$-values greater than 1. The weighted average PDFs for Sichuan-Yunnan and New Zealand show relatively small standard deviations, indicating strong consistency among the estimates from the four models and suggesting that the resulting $b^{\mathrm{pred}}$-values in these regions are precise and reliable. In the remaining four regions, the standard deviations are notably larger, reflecting greater variability across models. Our estimated $b^{\mathrm{pred}}$-values are generally consistent with previous studies that used traditional approaches such as least squares fitting, maximum likelihood estimation, or data-driven techniques over similar spatial and temporal domains. For example, $b \approx 1$ in the Beijing-Tianjin-Hebei region has been reported by Bi et al. (2024), Xiong et al. (2019), and Shen et al. (2024); similar values for the southeastern coast are noted in Cai et al. (2015). A $b$-value below 1 in the Sichuan-Yunnan region is consistent with findings from Xiong et al. (2020), Liu et al. (2024), and Zhang & Zhou (2016). In northern Xinjiang, $b < 1$ is supported by Zhang & Tang (2015), while $b > 1$ in California aligns with Kamer & Hiemer (2015) and Scholz (2015). Similarly, elevated $b$-values in New Zealand are consistent with Lu (2016, 2017).

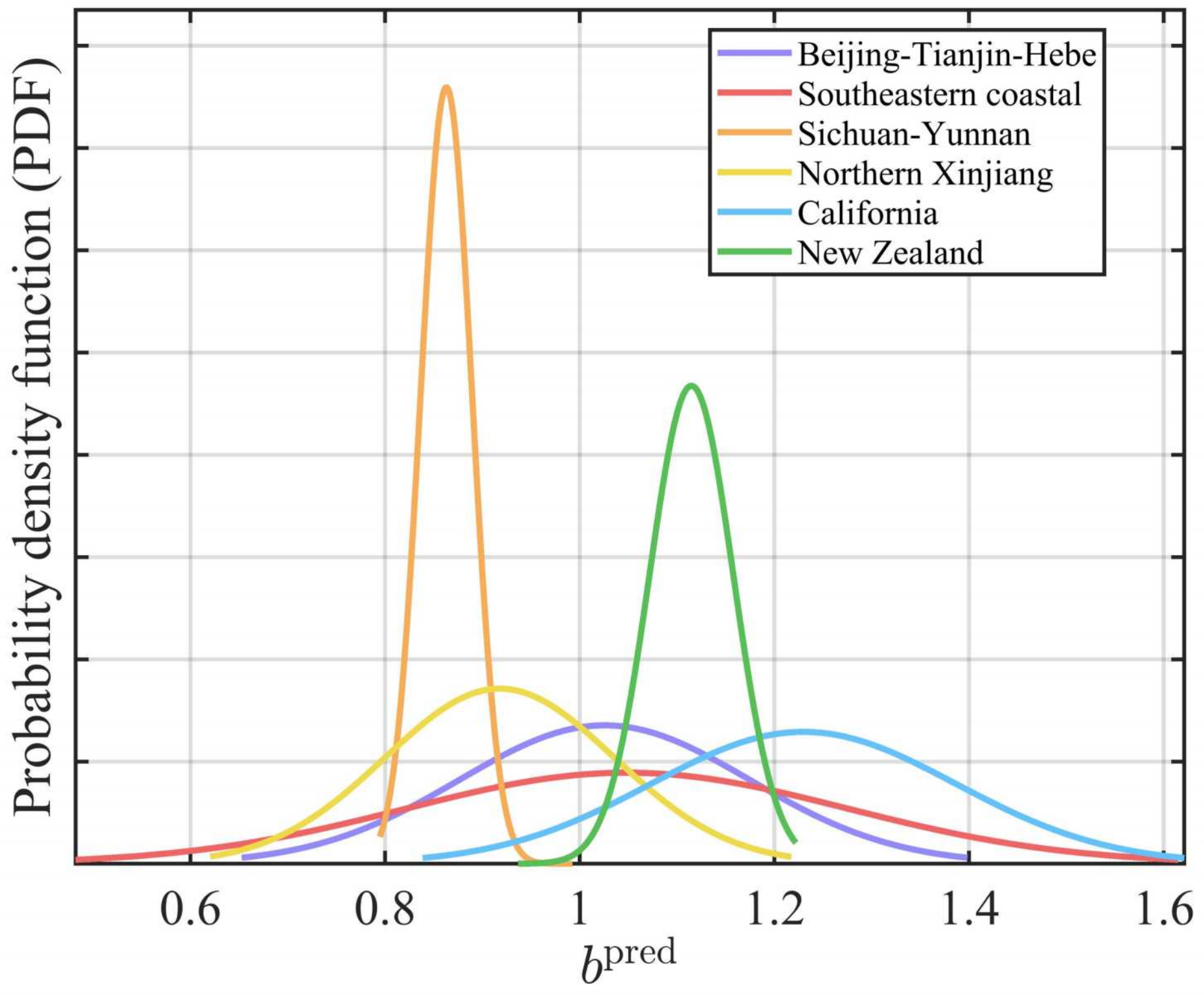

**Figure 3.** Weighted average of the $b^{\text{pred}}$-value probability density functions (PDFs) from the four augmented Gutenberg-Richter models for the six study regions. Details of the weighting procedure are described in Section 5.3. The individual PDFs of the estimated $b^{\text{pred}}$-value are provided in Figure S3 and Table 1.

## 6. Conclusion

We have proposed an augmented Gutenberg-Richter law incorporating four candidate gReLU($x$) functions designed specifically to handle catalog incompleteness. This new model integrates a probabilistic description of catalog completeness with the classical Gutenberg-Richter law for characterizing earthquake magnitude distributions. We evaluated the four augmented Gutenberg-Richter models (GR-SSReLU, GR-BSReLU, GR-COReLU, and GR-AEReLU) using synthetic earthquake catalogs—examining both correct model specification (where the fitting model matches the data-generating model) and model mis-specification scenarios—as well as observed earthquake catalogs collected since 2009 from six study regions. Model performance was systematically assessed based on parameter accuracy and uncertainty, goodness-of-fit metrics, and statistical $p$-values. Among the four models, GR-AEReLU emerged as the most robust choice, maintaining stable, reliable performance across varying scenarios. Thus, GR-AEReLU is recommended as the preferred model for general applications.

Our proposed augmented Gutenberg-Richter law offers a physically meaningful representation of the gradual transition between fully detected (complete) and partially detected (incomplete) portions of earthquake catalogs. Unlike traditional methods, which rely on a fixed, hard cutoff completeness magnitude $m_c$, this model interprets catalog completeness as a probabilistic concept characterized by a mean magnitude $m_c$ and a standard deviation $\sigma_c$. This formulation provides several advantages, especially in situations where the strict cut-off assumption inadequately reflects the complexities and inherent uncertainties associated with real seismic data. Specifically, the augmented Gutenberg-Richter law accounts for the decreasing probability of detecting smaller earthquakes below $m_c$, as well as the nonzero probability of missing larger events above $m_c$. The novel enhanced GR model offers distinct advantages: (1) it provides a unified framework that naturally integrates catalog completeness and magnitude-frequency distribution within a single, coherent formulation; (2) its parameters possess clear physical interpretations and intuitive statistical meanings; and (3) it effectively captures the nonlinear and asymmetric characteristics of earthquake detection probabilities, accurately reflecting real-world phenomena such as rapid detection probability decay and significant asymmetry at lower magnitudes, as commonly observed in empirical earthquake catalogs.

The findings summarised in Figure 3 challenge the long-held paradigm in seismology that the *b*-value is a universal constant equal to 1.0. This standard wisdom, which has underpinned earthquake statistics and hazard assessment for decades, appears to be an oversimplification that fails to capture the regional complexity of seismogenic processes. Our results demonstrate systematic and statistically significant deviations from $b = 1$ across different tectonic environments, with values ranging from below 0.9 in compressional settings like Sichuan-Yunnan and northern Xinjiang to above 1.1 in extensional or transform regions such as California and New Zealand.

The consistency of our physics-informed calibrations with traditional empirical estimates across multiple regions and studies suggests that these deviations are not merely artifacts of data quality or methodological choices, but rather reflect fundamental differences in the underlying physics of earthquake generation. The regional variability in *b*-values likely arises from variations in stress state, fault geometry, material heterogeneity, and thermal structure. These are factors that our augmented GR models explicitly incorporate through their general structure. This indicates that the assumption of a universal *b*-value may lead to biased estimates of seismic hazard, particularly in regions where the true *b*-value deviates significantly from unity.

Moving forward, the seismological community should reconsider the practice of defaulting to $b = 1$ in hazard assessments and instead adopt region-specific $b$-values that account for local tectonic conditions. Our physics-informed approach provides a framework for estimating these values even in data-sparse regions, offering a path toward more accurate and physically consistent earthquake statistics. This paradigm shift from a universal constant to regionally variable $b$-values represents a first step towards a fundamental advancement in our understanding of earthquake occurrence and has important implications for seismic hazard assessment worldwide.

Finally, by challenging the traditional assumption of absolute completeness above a fixed magnitude threshold, the augmented Gutenberg-Richter law provides a flexible, probabilistic framework capable of bridging conventional cutoff-based models and probabilistic completeness approaches. This flexibility is particularly advantageous for applications requiring detailed characterizations of data completeness. For example, our model can serve as a robust null hypothesis for identifying "dragon-king" earthquakes—extreme events that stand apart from typical frequency-magnitude relationships—without necessitating the specification of arbitrary truncation points (e.g., Li et al., 2025a; Li et al., 2025b). Moreover, integrating the augmented Gutenberg-Richter model into statistical seismicity frameworks such as the Epidemic-Type Aftershock Sequence (ETAS) model (e.g., Ogata, 1988, 1998, 2024; Ogata & Zhuang, 2006; Nandan et al., 2021; 2022) will allow catalog quality assessment to be jointly inferred through parameters $m_c$ and $\sigma_c$. Including previously discarded small earthquakes—appropriately weighted through these two parameters—can enhance the quality of statistical estimations of ETAS model parameters, such as productivity and branching ratios (Li et al., 2025c), and even help estimate the minimum magnitude capable of triggering aftershocks ($m_0$; Sornette & Werner, 2005; Zaccagnino et al., 2025). Overall, this innovative and practical augmented Gutenberg-Richter framework holds substantial promise for advancing our understanding of seismicity and improving the performance of statistical-physical earthquake models, thereby contributing to improved earthquake forecasting.

## Acknowledgments

The authors would like to thank Yang Zang from the China Earthquake Networks Center for his assistance in accessing the Chinese earthquake catalog data, as well as Prof. Jiancang Zhuang, and Prof. Zhongliang Wu for their valuable suggestions. This work is partially supported by the Guangdong Basic and Applied Basic Research Foundation (Grant No. 2024A1515011568), Shenzhen Science and Technology Innovation Commission (Grant no. GJHZ20210705141805017), and the Center for Computational Science and Engineering at Southern University of Science and Technology.

## Open Research

The California catalog is sourced from the Advanced National Seismic System (ANSS) Comprehensive Earthquake Catalog (ComCat; U.S. Geological Survey, 2017), accessible at https://earthquake.usgs.gov/data/comcat/ (last accessed: December 18, 2023). For the catalog in New Zealand, the data are obtained from the GeoNet Earthquake Catalog of New Zealand (GNS Science, 2022), accessible at https://quakesearch.geonet.org.nz/ (last accessed: December 18, 2023). We acknowledge the New Zealand GeoNet programme and its sponsors EQC, GNS Science, LINZ, NEMA and MBIE for providing data used in the present study. The data for the four study regions in China are acquired from the China Earthquake Networks Center (CENC) through the internal link provided by the Earthquake Cataloging System at China Earthquake Administration, available at http://10.5.160.18/console/index.action (last accessed: December 18, 2023), with a Digital Object Identifier (DOI) of 10.11998/SeisDmc/SN. This data is not publicly available; it can be requested from the CENC. The earthquake catalog for the four study regions in China, covering events with magnitudes ≥ 0 from 2009 to 2024 (excluding time and location information), is available in the "Example" folder at https://github.com/lijiawei098/GR-gReLU. The MATLAB codes for BSReLU, and AEReLU developed in this study are also openly accessible on the same GitHub repository.

## References

Akaike, H. (2011). Akaike's information criterion. International encyclopedia of statistical science, 25-25.

Aki, K. (1987). Magnitude-frequency relation for small earthquakes: A clue to the origin of *f*max of large earthquakes. Journal of Geophysical Research, 92(B2), 1349–1355. https://doi.org/10.1029/JB092iB02p01349

Bak, P., & Tang, C. (1989). Earthquakes as a self-organized critical phenomenon. Journal of Geophysical Research, 94(B11), 15,635–15,637. https://doi.org/10.1029/JB094iB11p15635

Black, F., & Scholes, M. (1973). The pricing of options and corporate liabilities. Journal of Political Economy, 81(3), 637–654. https://doi.org/10.1086/260062

Bouchaud, J.-P., & Sornette, D. (1994). The Black-Scholes option pricing problem in mathematical finance: Generalization and extensions for a large class of stochastic processes. Journal de Physique I, 4(6), 863–881. https://doi.org/10.1051/jp1:1994233

da Silva, S. L. E. (2021). $\kappa$-generalised Gutenberg-Richter law and the self-similarity of earthquakes. Chaos, Solitons & Fractals, 143, 110622. https://doi.org/10.1016/j.chaos.2020.110622

Darooneh, A. H., & Mehri, A. (2010). A nonextensive modification of the Gutenberg-Richter law: $q$-stretched exponential form. Physica A: Statistical

Mechanics and its Applications, 389(3), 509–514. https://doi.org/10.1016/j.physa.2009.09.024

Dayan, P., & Abbott, L. F. (2001). Theoretical Neuroscience: Computational and Mathematical Modeling of Neural Systems. MIT Press.

Ding, H., Zhou, Y., Ge, Z., Taymaz, T., Ghosh, A., Xu, H., et al. (2023). High-resolution seismicity imaging and early aftershock migration of the 2023 Kahramanmaraş (SE Türkiye) $M_W$ 7.9 & 7.8 earthquake doublet. Earthquake Science, 36(6), 417–432. https://doi.org/10.29382/eqs-2023-0032

Feng, Y., Mignan, A., Sornette, D., & Li, J. (2022). Hierarchical Bayesian modeling for improved high-resolution mapping of the completeness magnitude of earthquake catalogs. Seismological Research Letters, 93(4), 2126-2137.

García-Hernández, R., D'Auria, L., Barrancos, J., & Padilla, G. D. (2019). On the functional expression of frequency-magnitude distributions: A comprehensive statistical examination. Bulletin of the Seismological Society of America, 109(1), 482–486. https://doi.org/10.1785/0120180150

Geller, R. J., Jackson, D. D., Kagan, Y. Y., & Mulargia, F. (1997). Earthquakes cannot be predicted. Science, 275(5306), 1616–1617. https://doi.org/10.1126/science.275.5306.1616

GNS Science. (2022). GeoNet Aotearoa New Zealand Stations Metadata Repository [Dataset]. GNS Science, GeoNet. https://doi.org/10.21420/0S8P-TZ38.

Gutenberg, B., & Richter, C. F. (1944). Frequency of earthquakes in California. Bulletin of the Seismological Society of America, 34(4), 185–188.

Gutenberg, B., & Richter, C. F. (1954). Seismicity of the Earth and Associated Phenomena (2nd ed.). Princeton University Press.

Jia, K., & Zhou, S. (2024). Machine learning applications in seismology. Applied Sciences, 14(17), 7857. https://doi.org/10.3390/app14177857

Kagan, Y. Y. (1997). Earthquake size distribution and earthquake insurance. Communications in Statistics: Stochastic Models, 13(4), 775–797. https://doi.org/10.1080/15326349708807407

Kagan, Y.Y. (2002a), Seismic moment distribution revisited: I. Statistical results, Geophysical Journal International 148 (3), 520-541.

Kagan, Y.Y. (2002b), Seismic moment distribution revisited: II. Moment conservation principle, Geophysical Journal International 149 (3), 731-754.

Kijko, A., & Smit, A. (2017). Estimation of the frequency-magnitude Gutenberg-Richter *b*-value without making assumptions on levels of completeness. Seismological Research Letters, 88(2A), 311–318. https://doi.org/10.1785/0220160158

Klügel, J. U., Mualchin, L., & Panza, G. F. (2006). A scenario-based procedure for seismic risk analysis. Engineering Geology, 88(1–2), 1–22. https://doi.org/10.1016/j.enggeo.2006.08.003

Kong, Q., Trugman, D. T., Ross, Z. E., Bianco, M. J., Meade, B. J., & Gerstoft, P. (2019). Machine learning in seismology: Turning data into insights. Seismological Research Letters, 90(1), 3–14. https://doi.org/10.1785/0220180251

Kubo, H., Naoi, M., & Kano, M. (2024). Recent advances in earthquake seismology using machine learning. Earth, Planets and Space, 76(1), 36. https://doi.org/10.1186/s40623-024-01866-7

Li, J., Hao, M., & Cui, Z. (2024). A high-resolution aftershock catalog for the 2014 $M_S$ 6.5 Ludian (China) earthquake using deep learning methods. Applied Sciences, 14(5), 1997. https://doi.org/10.3390/app14051997

Li, J., Mignan, A., Sornette, D., & Feng, Y. (2023). Predicting the future performance of the planned seismic network in Chinese mainland. Seismological Research Letters, 94(5), 2698–2711. https://doi.org/10.1785/0220230102

Li, J., Sornette, D., Wu, Z., & Li, H. (2025a). New horizon in the statistical physics of earthquakes: Dragon-king theory and dragon-king earthquakes. Reviews of Geophysics and Planetary Physics, 56(3), 225–242. https://doi.org/10.19975/j.dqyxx.2024-033

Li, J., and Sornette, D. (2025b). Are Haicheng and Tangshan Earthquakes Dragon-Kings?. Journal of Geophysical Research: Solid Earth (submitted). Preprint available at http://arxiv.org/abs/2504.21310

Li, J., Sornette, D., Wu, Z., Zhuang, J., & Jiang, C. (2025c). Revisiting seismicity criticality: A new framework for bias correction of statistical seismology model calibrations. Journal of Geophysical Research: Solid Earth, 130, e2024JB029337.

Liu, Q., & Furber, S. (2016). Noisy softplus: A biology inspired activation function. In Proceedings of the 23rd International Conference on Neural Information Processing (pp. 405–412). Springer. https://doi.org/10.1007/978-3-319-46681-1_49

Mignan, A.(2012) Functional shape of the earthquake frequency magnitude distribution and completeness magnitude, J. Geophys. Res., 117(B8).

Mignan, A., & Woessner, J. (2012). Estimating the magnitude of completeness for earthquake catalogs. Community Online Resource for Statistical Seismicity Analysis. https://doi.org/10.5078/corssa-00180805

Mignan, A., Jiang, C., Zechar, J. D., Wiemer, S., Wu, Z., & Huang, Z. (2013). Completeness of the Mainland China earthquake catalog and implications for the setup of the China Earthquake Forecast Testing Center. Bulletin of the Seismological Society of America, 103(2A), 845–859. https://doi.org/10.1785/0120120052

Mizrahi, L., Dallo, I., van der Elst, N. J., Christophersen, A., Spassiani, I., Werner, M. J., ... & Wiemer, S. (2024). Developing, testing, and communicating earthquake forecasts: Current practices and future directions. Reviews of Geophysics, 62(3), e2023RG000823.

Mousavi, S. M., & Beroza, G. C. (2022). Deep-learning seismology. Science, 377(6607), eabm4470.

Mousavi, S. M., & Beroza, G. C. (2023). Machine learning in earthquake seismology. Annual Review of Earth and Planetary Sciences, 51(1), 105-129.

Nair, V., & Hinton, G. E. (2010). Rectified linear units improve restricted Boltzmann machines. In Proceedings of the 27th International Conference on Machine

Learning (pp. 807–814).
Nandan, S., Ram, S. K., Ouillon, G., & Sornette, D. (2021). Is seismicity operating at a critical point? Physical Review Letters, 126(12), 128501. https://doi.org/10.1103/PhysRevLett.126.128501
Nandan, S., Ouillon, G., & Sornette, D. (2022). Are large earthquakes preferentially triggered by other large events? Journal of Geophysical Research, 127(8), e2022JB024380. https://doi.org/10.1029/2022JB024380
Ogata, Y. (1988). Statistical models for earthquake occurrences and residual analysis for point processes. Journal of the American Statistical Association, 83(401), 9–27. https://doi.org/10.1080/01621459.1988.10478560
Ogata, Y. (1998). Space-time point-process models for earthquake occurrences. Annals of the Institute of Statistical Mathematics, 50(2), 379–402. https://doi.org/10.1023/A:1003403601725
Ogata, Y., & Katsura, K. (1993). Analysis of temporal and spatial heterogeneity of magnitude frequency distribution inferred from earthquake catalogues. Geophysical Journal International, 113(3), 727–738. https://doi.org/10.1111/j.1365-246X.1993.tb04662.x
Ogata, Y., & Katsura, K. (2006), Immediate and updated forecasting of aftershock hazard. Geophysical Research Letters, 33(10). doi:10.1029/2006GL025888
Ogata, Y., & Zhuang, J. (2006). Space-time ETAS models and an improved extension. Tectonophysics, 413(1–2), 13–23. https://doi.org/10.1016/j.tecto.2005.10.016
Ogata, Y. (2024). How the ETAS models were created, used, and evolved – Personal views and perspectives. Annals of Geophysics, 67(4), S428. https://doi.org/10.4401/ag-9153
Petersen, T., Gledhill, K., Chadwick, M., Gale, N. H., & Ristau, J. (2011). The New Zealand National Seismograph Network. Seismological Research Letters, 82(1), 9–20. https://doi.org/10.1785/gssrl.82.1.9
Ringdal, F. (1975), On the estimation of seismic detection thresholds, Bull. Seismol. Soc. Am., 65, 1631–1642.
Rydelek, P. A., & Sacks, I. S. (1989). Testing the completeness of earthquake catalogues and the hypothesis of self-similarity. Nature, 337(6204), 251–253. https://doi.org/10.1038/337251a0
Schorlemmer, D., & Woessner, J. (2008). Probability of detecting an earthquake. Bulletin of the Seismological Society of America, 98(5), 2103–2117. https://doi.org/10.1785/0120070105
Silva, R., Franca, G. S., Vilar, C. S., & Alcaniz, J. S. (2006). Nonextensive models for earthquakes. Physical Review E, 73(2), 026102. https://doi.org/10.1103/PhysRevE.73.026102
Sornette, A., & Sornette, D. (1989). Self-organized criticality and earthquakes. Europhysics Letters, 9(3), 197–202. https://doi.org/10.1209/0295-5075/9/3/002
Sornette, D., Knopoff, L., Kagan, Y. Y., & Vanneste, C. (1996). Rank-ordering statistics of extreme events: Application to the distribution of large earthquakes. Journal of Geophysical Research: Solid Earth, 101(B6), 13883–13893. https://doi.org/10.1029/96JB00177

Sornette, D., & Sornette, A. (1999). General theory of the modified Gutenberg-Richter law for large seismic moments. Bulletin of the Seismological Society of America, 89(4), 1121–1130.

Sornette, D., & Werner, M. J. (2005). Constraints on the size of the smallest triggering earthquake from the epidemic-type aftershock sequence model, Båth's law, and observed aftershock sequences. Journal of Geophysical Research: Solid Earth, 110(B8), B08304. https://doi.org/10.1029/2005JB003535

Sotolongo-Costa, O., & Posadas, A. (2004). Fragment-asperity interaction model for earthquakes. Physical Review Letters, 92(4), 048501. https://doi.org/10.1103/PhysRevLett.92.048501

Tormann, T., Wiemer, S., & Mignan, A. (2014). Systematic survey of high-resolution b value imaging along Californian faults: Inference on asperities. Journal of Geophysical Research, 119(3), 2029–2054. https://doi.org/10.1002/2013JB010867

U.S. Geological Survey. (2017). Advanced National Seismic System (ANSS) comprehensive catalog of earthquake events and products. https://doi.org/10.5066/F7MS3QZH

van der Elst, N. J. (2021). B-Positive: A robust estimator of aftershock magnitude distribution in transiently incomplete catalogs. Journal of Geophysical Research: Solid Earth, 126(2), e2020JB021027. https://doi.org/10.1029/2020JB021027

Wang, X., Li, J., Feng, A., & Sornette, D. (2025). Estimating magnitude completeness in earthquake catalogs: A comparative study of catalog-based methods. Manuscript submitted for publication. Preprint available at https://doi.org/10.48550/arXiv.2502.17838

Wiemann, P. F., Kneib, T., & Hambuckers, J. (2021). Using the softplus function to construct alternative link functions in generalized linear models and beyond. Statistical Papers, 62(2), 1–26. https://doi.org/10.1007/s00362-020-01211-3

Woessner, J., & Wiemer, S. (2005). Assessing the quality of earthquake catalogues: Estimating the magnitude of completeness and its uncertainty. Bulletin of the Seismological Society of America, 95(2), 684–698. https://doi.org/10.1785/0120040007

Zaccagnino, D., Li, J., & Sornette, D. (2025). The UV-divergence problem in statistical seismology: Insights from an ETAS model with smoothed minimum triggering magnitude. In Abstracts of the 43rd National Conference of the Gruppo Nazionale di Geofisica della Terra Solida (GNGTS), February 11–14, 2025, Bologna, Italy.

Zhou, Y., Ghosh, A., Fang, L., Yue, H., Zhou, S., & Su, Y. (2021). A high-resolution seismic catalog for the 2021 MS6.4/MW6.1 Yangbi earthquake sequence, Yunnan, China: Application of AI picker and matched filter. Earthquake Science, 34(5), 390–398. https://doi.org/10.29382/eqs-2021-0031

*Support Information (Texts) to*

# Unifying the Gutenberg-Richter Law with Probabilistic Catalog Completeness

Jiawei Li[1], Xinyi Wang[1,2], and Didier Sornette[1]

1 Institute of Risk Analysis, Prediction and Management (Risks-X), Academy for Advanced Interdisciplinary Studies, Southern University of Science and Technology (SUSTech), Shenzhen, China.

2 School of Computer Science, Chengdu University of Information Technology (CUIT), Chengdu, China.

Corresponding authors: Didier Sornette (didier@sustech.edu.cn); Jiawei Li (lijw@cea-igp.ac.cn)

# Jiawei Li and Xinyi Wang contributed equally to this work

**Text S1.** Fitting algorithm for the augmented Gutenberg-Richter models

To enhance both computational efficiency and numerical robustness—particularly to mitigate the impact of tail-region residual perturbations (Figure S11)—we employ a weighted residual nonlinear least squares fitting method. The weighting scheme is based on the square root of the normalized complementary cumulative distribution function, i.e., $\sqrt{\log_{10}\mathrm{CCDF}(m)}$, which assigns greater weight to the more stable, high-frequency regions without allowing extreme values to dominate the fit. This strategy strikes a balance between emphasizing reliable regions of the distribution and accounting for the heteroscedastic nature of frequency-magnitude distribution, where variance decreases as frequency increases. To further avoid local minima, five parallel optimizations are performed using randomly perturbed initial parameter sets, and the solution with the lowest weighted residual is retained.

**Text S2.** Performance on simulated catalog

Tables S1(a) to S4(a) show, overall, as the sample size parameter $k$ increases, the fitted parameters $\theta^{\text{pred}} = \{a^{\text{pred}}, b^{\text{pred}}, m^{\text{pred}}{}_{\text{c}}, \sigma_{\text{c}}^{\text{pred}}, \beta^{\text{pred}}\}$ progressively converge toward their predefined true values. Simultaneously, the uncertainty associated with the parameter estimates—quantified by the standard deviations from 200 independently sampled datasets—tends to decrease with increasing $k$.

To provide further insight, the statistics are broken down into two categories for the simulated catalog: (1) correct model specification—cases where each model is fitted to data generated by itself, corresponding to the light red-shaded rows in Tables S1(a) to S4(a), and (2) model mis-specification — cases where a model is fitted to data generated by a different model, corresponding to the light blue-shaded rows in Tables S1(a) to S4(a). In the cases of correct model specification, all models exhibit favorable performances (1st or 2nd place).

In the cases of model mis-specification, all models exhibit a decrease in performance compared to correct model specification. Among these models, GR-BSReLU and GR-COReLU generally yield weaker results, often ranked as poor (light blue) or worst (dark blue) across all evaluation metrics. GR-SSReLU exhibits intermediate performance. GR-AEReLU is most often ranked first or second, respectively, with no occurrences of being ranked as worst. These results indicate the superior adaptability of GR-AEReLU, especially under conditions of model mis-specification. This can be attributed to the greater flexibility and smoothness of GR-AEReLU; notably, GR-COReLU is a special case of GR-AEReLU with the shape parameter $\beta = 0$, limiting its adaptability.

**Text S3.** Performance on empirical catalog

Table 1 shows the fitted parameters of the augmented Gutenberg-Richter models applied to empirical earthquake catalogs from six study regions, including four regions in China (Beijing-Tianjin-Hebei, the southeastern coast, Sichuan-Yunnan, and northern Xinjiang), as well as California and New Zealand. The values in Table 1 are the means and standard deviations derived from 200 bootstrap iterations. Each bootstrap iteration involves resampling the empirical catalog with replacement to generate a dataset of equal size, resulting in 200 distinct catalog realizations.

Based on the standard deviations of $m_c^{pred}$, $a^{pred}$, $b^{pred}$, and $\sigma_c^{pred}$ (Table 1; Figures S2 to S5), GR-AEReLU, GR-SSReLU and GR-COReLU generally exhibit small parameter estimation uncertainties. The GR-BSReLU model consistently shows the highest uncertainty, with standard deviations typically an order of magnitude larger than those of the other three models. In terms the width of the transition region from complete to incomplete quantified by $\sigma_c^{pred}$, GR-AEReLU consistently achieves values ranging 0.3 to 0.4 across all six study regions (Table 1; Figure S5). GR-SSReLU and GR-COReLU gives similar results, while GR-BSReLU yields larger estimates of $\sigma_c^{pred}$. However, when compared to the other models, these elevated values appear to reflect not a genuine structure of the seismic catalog, but rather a reduced ability of GR-BSReLU to accurately and precisely model the censored empirical distribution of earthquake magnitudes.

Table S5(a) also reports the Rmse, R-Square, SSE, and AIC performance metrics for the augmented Gutenberg-Richter models, while Table S5(b) provides the corresponding rankings of the augmented Gutenberg-Richter models based on these metrics. These rankings are also visually summarized through pie charts in Figure 2. It can be observed that GR-BSReLU and GR-COReLU tend to exhibit polarized performance, frequently ranking either as the best or the worst across evaluations. In contrast, GR-SSReLU and GR-AEReLU tend to occupy intermediate performance ranks, with GR-AEReLU standing out by attaining best or good rankings significantly more often than the other three models.

Table S5 also report the *p*-values and the corresponding model rankings. The *p*-values were computed following these steps: (1) Fit an empirical catalog to obtain the best-fitting model (Figures S7 to S10) and calculate the Kolmogorov-Smirnov (KS) statistic of the observed catalog relative to the best-fitting model, denoted as $KS_{obs}$; (2) Generate 10,000 sets of simulated catalogs using the best-fitting model, and compute the KS statistic for each, denoted as $KS^i_{sim}$ ($i$ = 1, 2, ..., 10,000); (3) Calculate the *p*-value as the proportion of $KS^i_{sim}$ that are larger than $KS_{obs}$. To simulate randomness within the synthetic earthquake catalogs, a uniform random noise uniformly sampled in the interval [-0.5, 0.5] was added to each synthetic magnitude. According to the *p*-value rankings, GR-SSReLU and GR-AEReLU perform the best, being ranked as best or good in 100% of cases. GR-BSReLU performs comparably and fall in the middle range, each being ranked as best in about 67% of cases. GR-COReLU performs the worst, receiving a best ranking in 33.3% of the cases but a worst ranking in 33.3% as well.

**Text S4.** GR-AEReLU results on empirical catalog

The Rmse values for GR-AEReLU vary between approximately $4.0 \times 10^2$ and $6.0 \times 10^3$ across the six regions. The highest Rmse is observed in Sichuan-Yunnan [$(5.79 \pm 0.34) \times 10^3$] and California [$(4.50 \pm 1.04) \times 10^3$], while the lowest occurs in northern Xinjiang [$(3.96 \pm 1.16) \times 10^2$] and Southeastern coastal [$(4.59 \pm 1.13) \times 10^2$]. The $R$-Square values are above 0.9985 in all regions, indicating excellent model fit. The SSE and AIC metrics, however, depend on dataset size and thus are not directly comparable across the six study areas. The GR-AEReLU model demonstrates consistently strong performance across diverse regions, with notably high $p$-values ($\geq$ 0.8942) indicating excellent fits for northern Xinjiang, California, southeastern coastal of China, and Sichuan-Yunnan regions. Moderate results in New Zealand ($p$-value = 0.5240) and Beijing-Tianjin-Hebei ($p$-value = 0.3050) highlight regional complexities, affirming the model's flexibility and robustness in capturing realistic earthquake completeness and magnitude distributions.

For six regions, the uncertainties in the estimated $a^{\text{pred}}$ and $b^{\text{pred}}$-values remain below 0.14 and 0.06, respectively. In terms of the mean $b^{\text{pred}}$-values, Sichuan-Yunnan (0.87 ± 0.01) yields values below 1.0, while northern Xinjiang (0.92 ± 0.06), the southeastern coast (0.96 ± 0.05) and Beijing-Tianjin-Hebei (0.97 ± 0.04) are closest to 1. In contrast, California (1.17 ± 0.04) and New Zealand (1.11 ± 0.02) exhibit mean $b^{\text{pred}}$-values greater than 1.

For $m_c^{\text{pred}}$, the uncertainty in its parameter estimates across all regions is below 0.16, and the uncertainty in $\sigma_c^{\text{pred}}$ is consistently less than 0.06, reflecting the robustness of these estimates. As outlined in Section 2, $\sigma_c$ quantifies the intrinsic smooth transition from incompleteness to completeness as magnitudes crossover from below to above $m_c$ and should not be confused with the standard error of $m_c^{\text{pred}}$ reported in Table 1. Based on the GR-AEReLU results, the estimated $m_c^{\text{pred}}$ and $\sigma_c^{\text{pred}}$ values are approximately 0.76 and 0.38 for Beijing-Tianjin-Hebei, 0.58 and 0.32 for the southeastern coast, 1.09 and 0.39 for Sichuan-Yunnan, 1.47 and 0.39 for northern Xinjiang, 0.96 and 0.41 for California, and 2.07 and 0.39 for New Zealand. As described in Section 5.1, we further computed completeness probabilities at $m = m_c^{\text{pred}} + \sigma_c^{\text{pred}}$, $m = m_c^{\text{pred}} + 2\sigma_c^{\text{pred}}$, and $m = m_c^{\text{pred}} + 3\sigma_c^{\text{pred}}$ for all six study regions. These results highlight the capacity of GR-AEReLU to provide not only stable parameter estimates but also a probabilistic description of catalog completeness, which can guide users in selecting magnitude thresholds based on the desired level of completeness.

*Support Information (Figures) to*

# Unifying the Gutenberg-Richter Law with Probabilistic Catalog Completeness

Jiawei Li[1], Xinyi Wang[1,2], and Didier Sornette[1]

1 Institute of Risk Analysis, Prediction and Management (Risks-X), Academy for Advanced Interdisciplinary Studies, Southern University of Science and Technology (SUSTech), Shenzhen, China.
2 School of Computer Science, Chengdu University of Information Technology (CUIT), Chengdu, China.

Corresponding authors: Didier Sornette (didier@sustech.edu.cn); Jiawei Li (lijw@cea-igp.ac.cn)

# Jiawei Li and Xinyi Wang contributed equally to this work

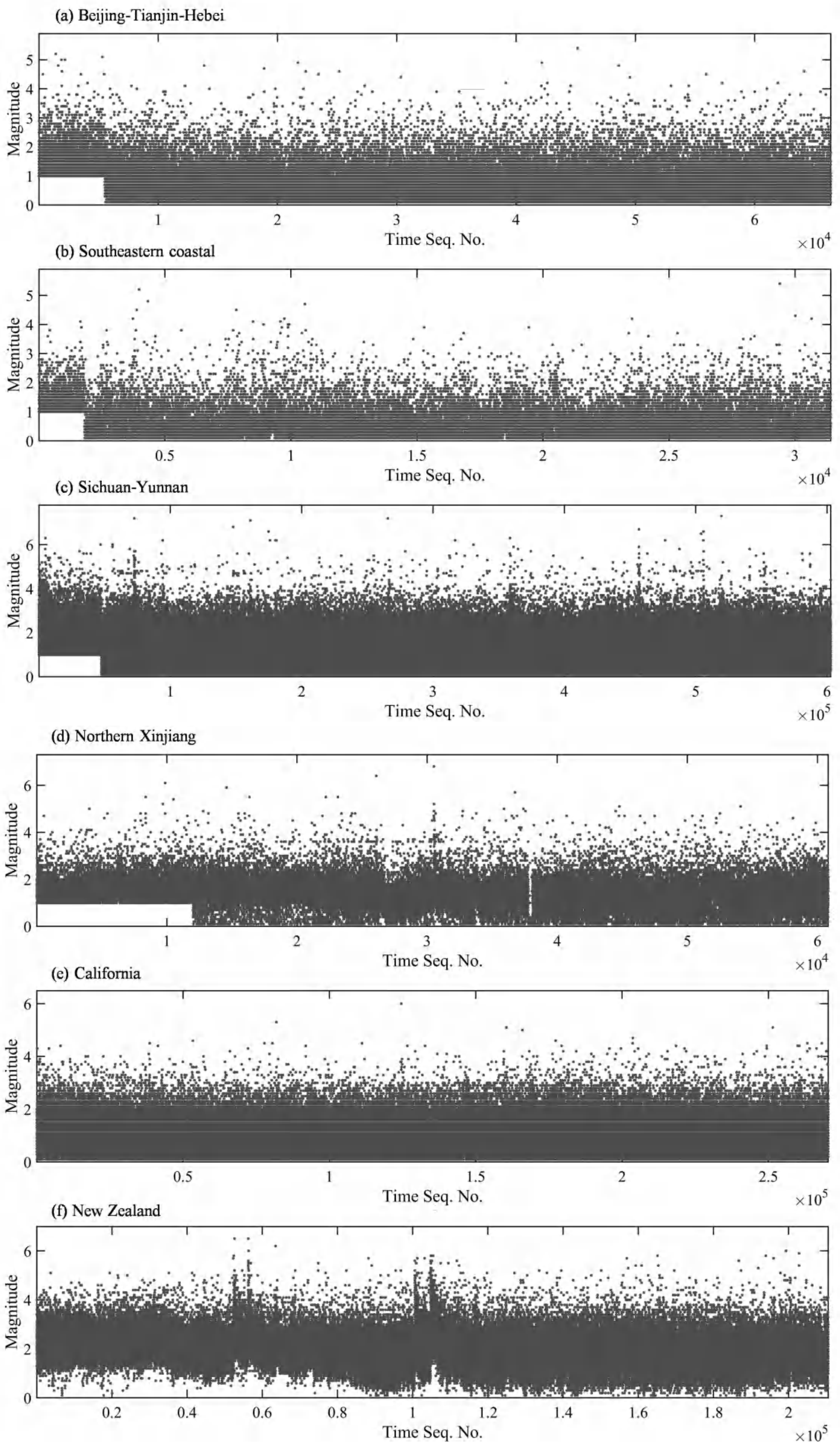

**Figure S1.** Magnitude-time sequence plot of the empirical earthquake catalogs from six study regions. The magnitudes are perturbed by adding random errors uniformly distributed within the interval [-0.05, 0.05].

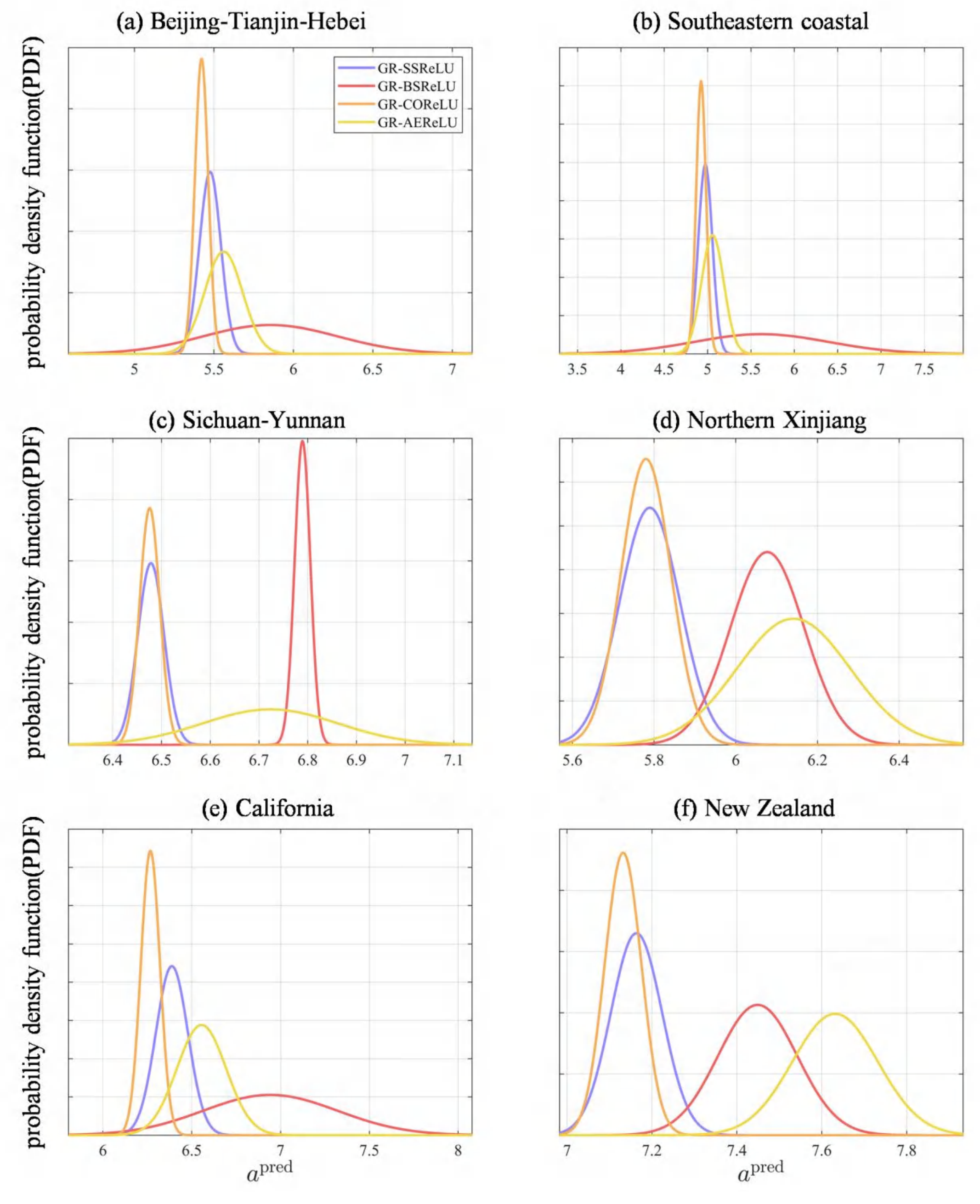

**Figure S2.** Probability density function (PDF) of the estimated $a^{\text{pred}}$ values obtained from 200 bootstrapping iterations using the augmented Gutenberg-Richter law with the four gReLU($x$) functions, applied to empirical earthquake catalogs from six study regions.

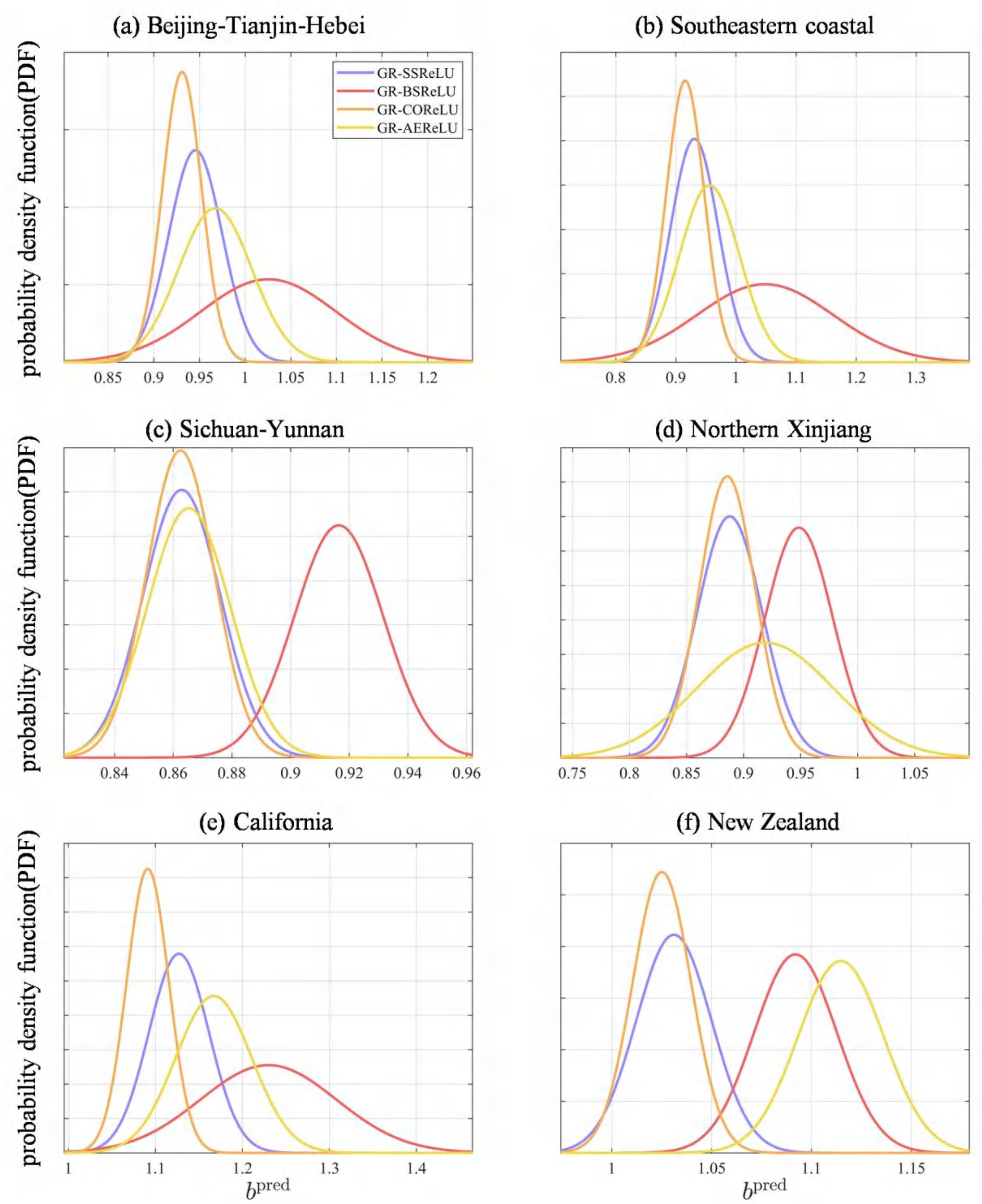

**Figure S3.** PDF of the estimated $b^{pred}$ values obtained from 200 bootstrapping iterations using the augmented Gutenberg-Richter law with the four gReLU($x$) functions, applied to empirical earthquake catalogs from six study regions.

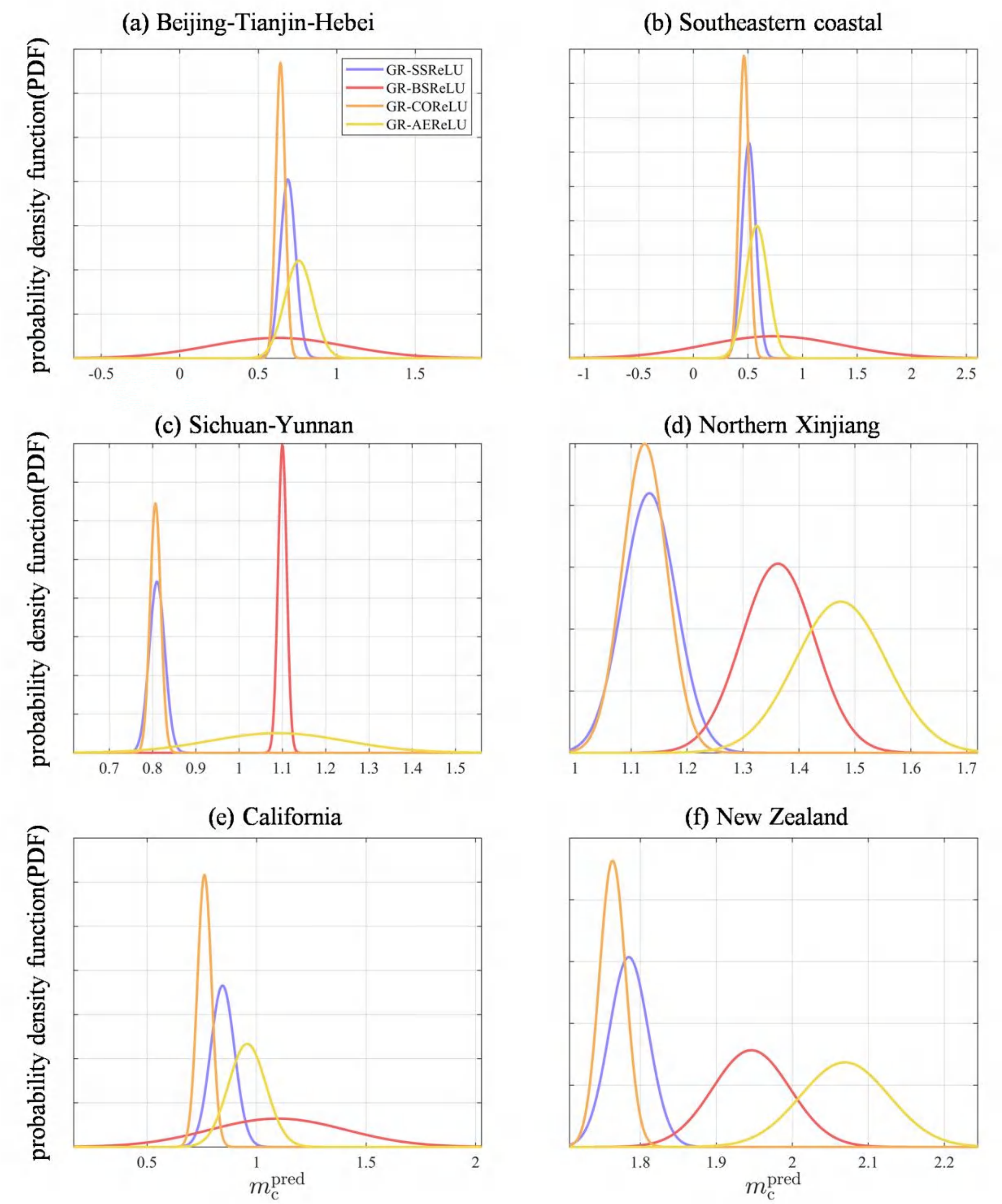

**Figure S4.** PDF of the estimated $m_c^{pred}$ values obtained from 200 bootstrapping iterations using the augmented Gutenberg-Richter law with the four gReLU($x$) functions, applied to observed earthquake catalogs from six study regions.

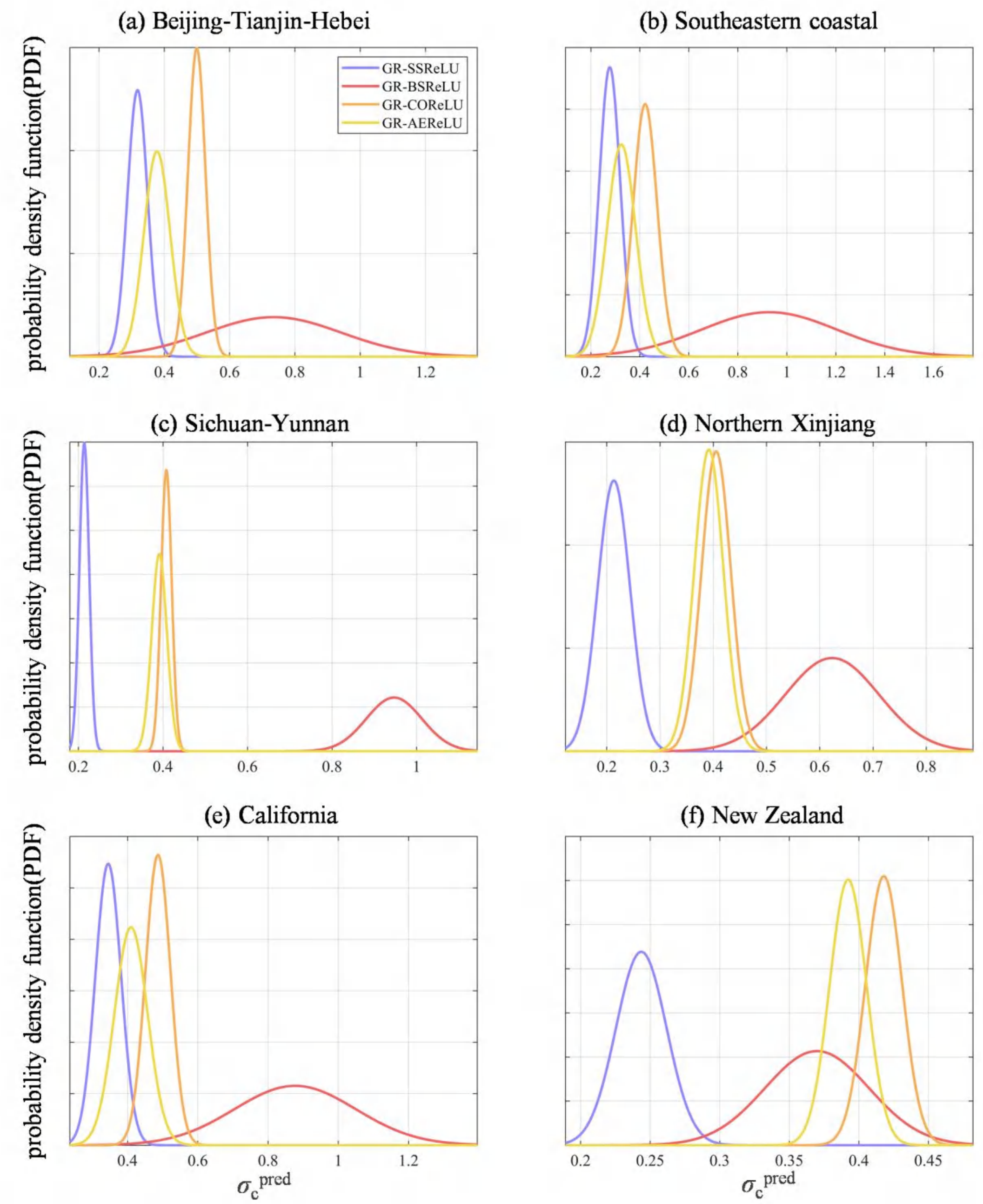

**Figure S5.** PDF of the estimated $\sigma_c^{pred}$ values obtained from 200 bootstrapping iterations using the augmented Gutenberg-Richter law with the four gReLU($x$) functions, applied to empirical earthquake catalogs from six study regions.

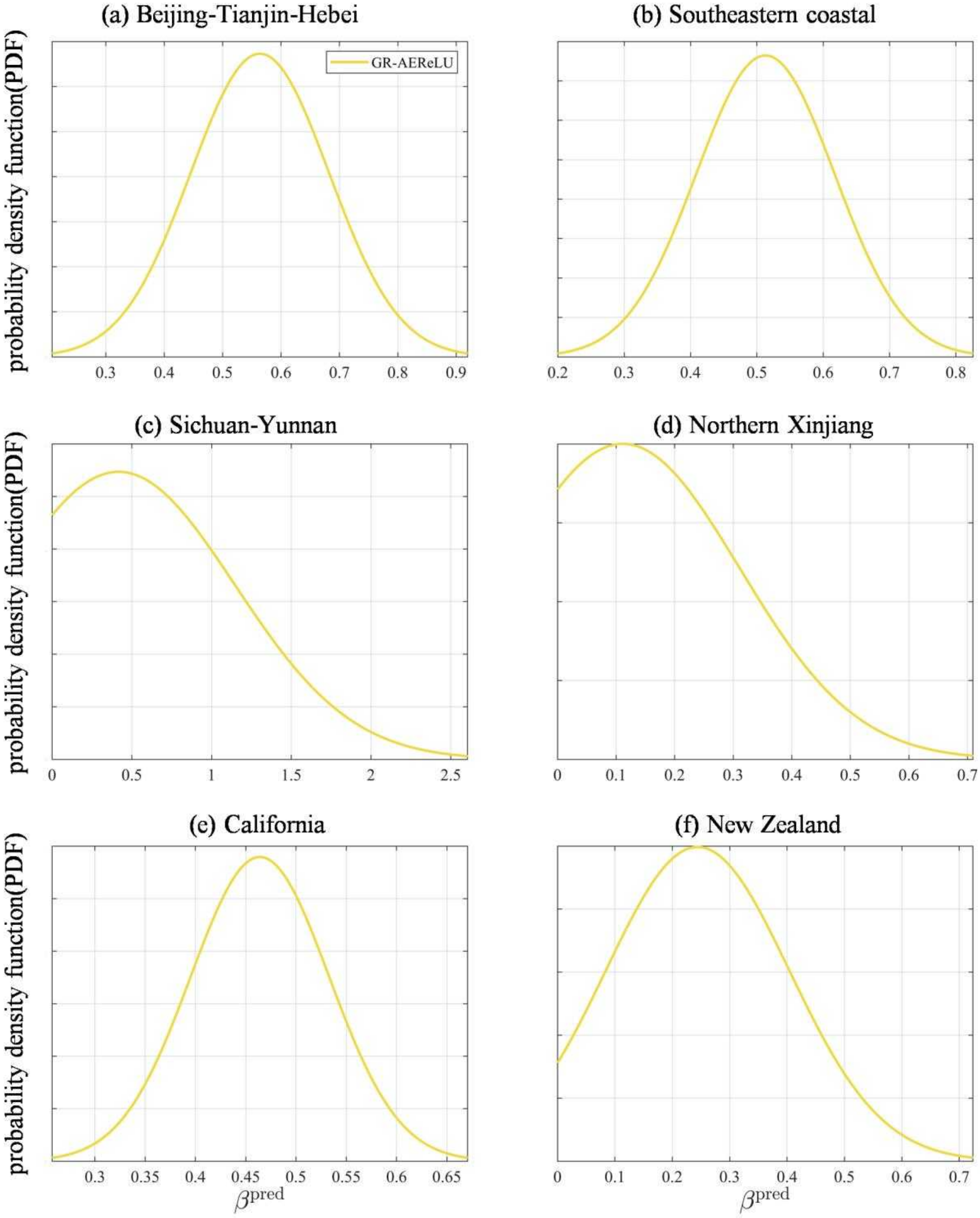

**Figure S6.** PDF of the estimated $\beta^{pred}$ values obtained from 200 bootstrapping iterations using the augmented Gutenberg-Richter law with the AEReLU($x$) function, applied to empirical earthquake catalogs from six study regions.

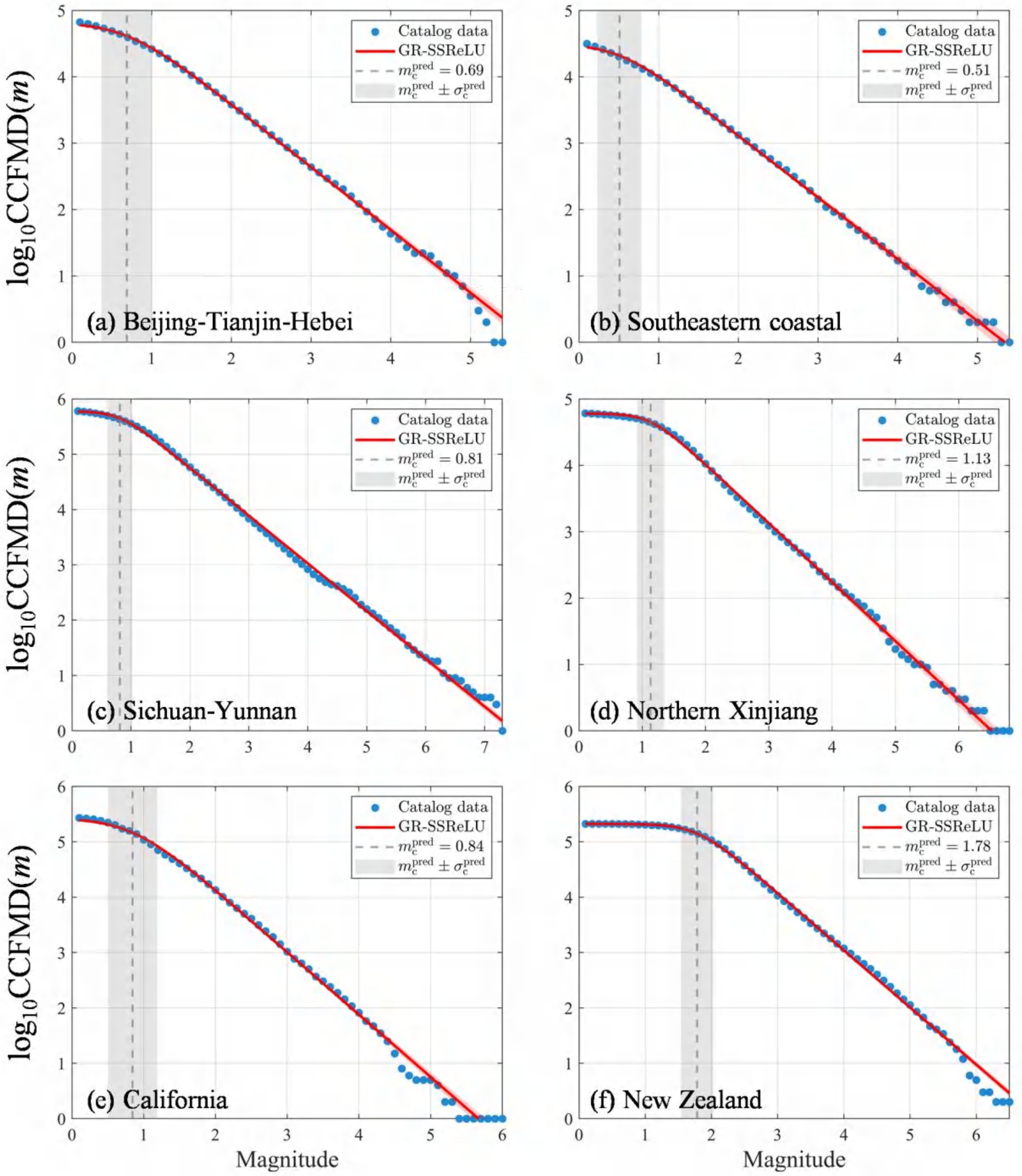

**Figure S7.** Complementary cumulative frequency-magnitude distribution (CCFMD) of the empirical earthquake catalogs from six study regions, along with the fitted CCFMD($m$) derived using the augmented Gutenberg-Richter law with the SSReLU($x$) function (GR-SSReLU). The red shading represents the estimated uncertainty, obtained from 200 bootstrapping iterations.

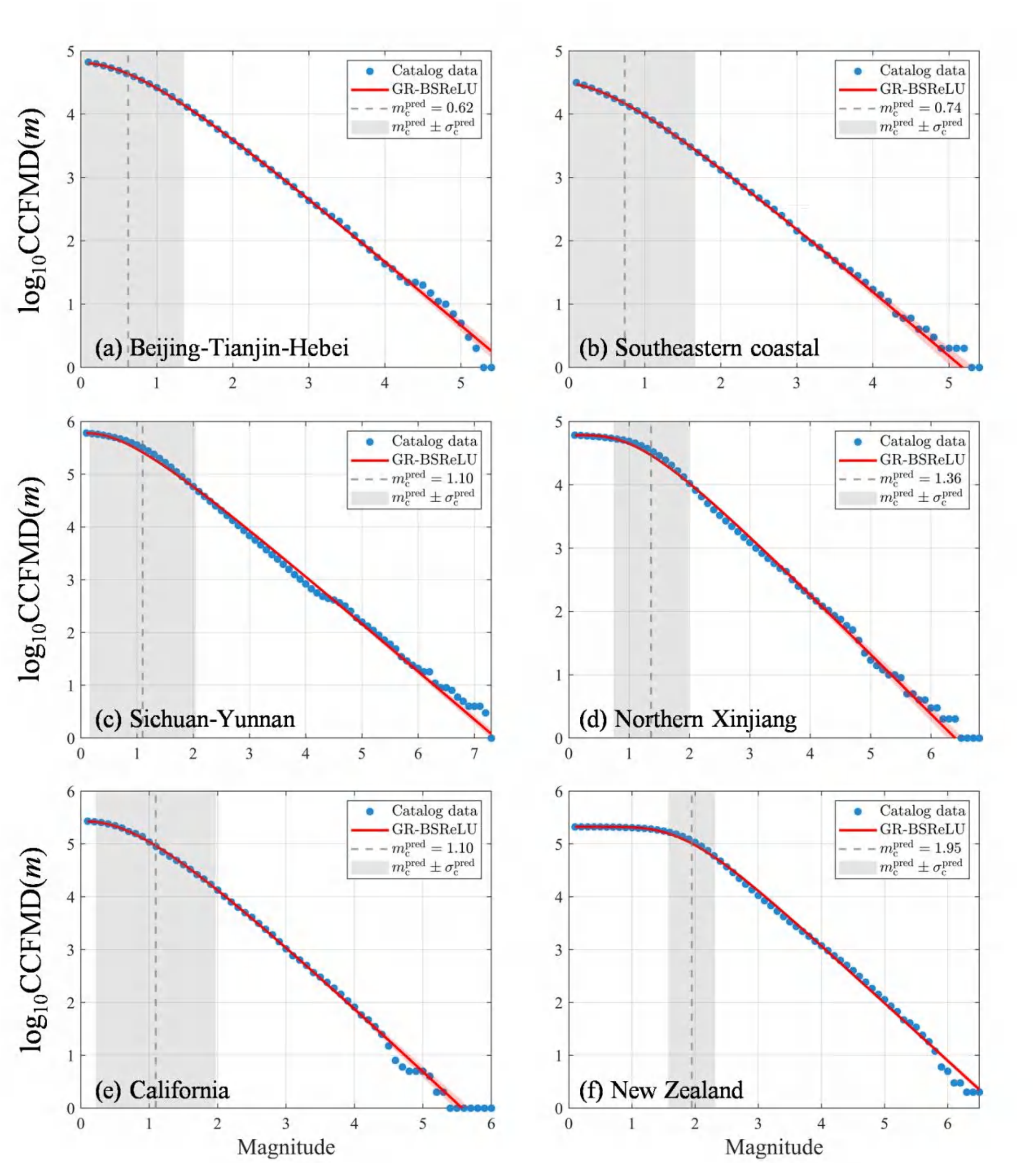

**Figure S8.** CCFMD of the empirical earthquake catalogs from six study regions, along with the fitted CCFMD($m$) derived using the augmented Gutenberg-Richter law with the BSReLU($x$) function. The red shading represents the estimated uncertainty, obtained from 200 bootstrapping iterations.

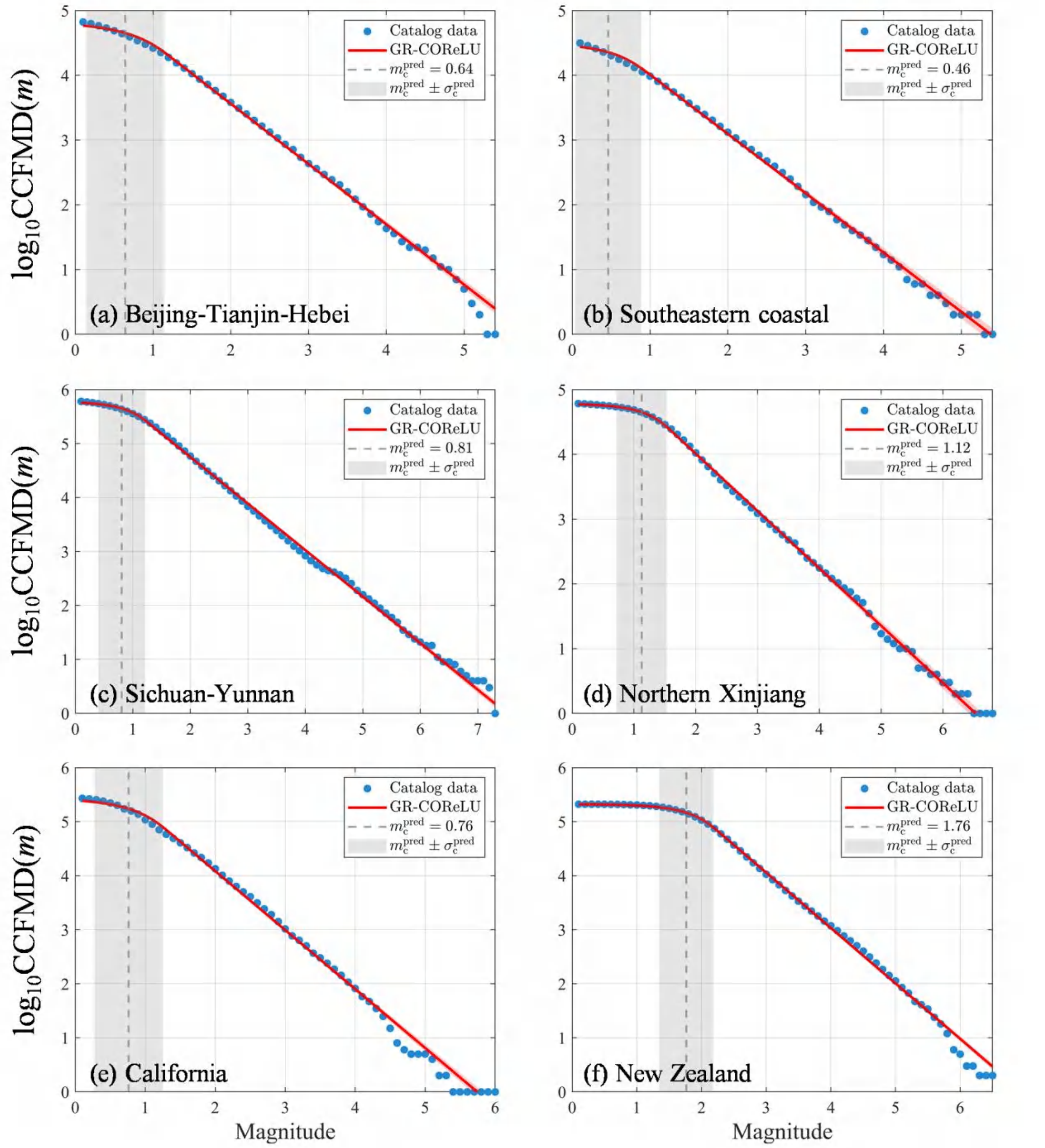

**Figure S9.** CCFMD of the empirical earthquake catalogs from six study regions, along with the fitted CCFMD($m$) derived using the augmented Gutenberg-Richter law with the COReLU($x$) function. The red shading represents the estimated uncertainty, obtained from 200 bootstrapping iterations.

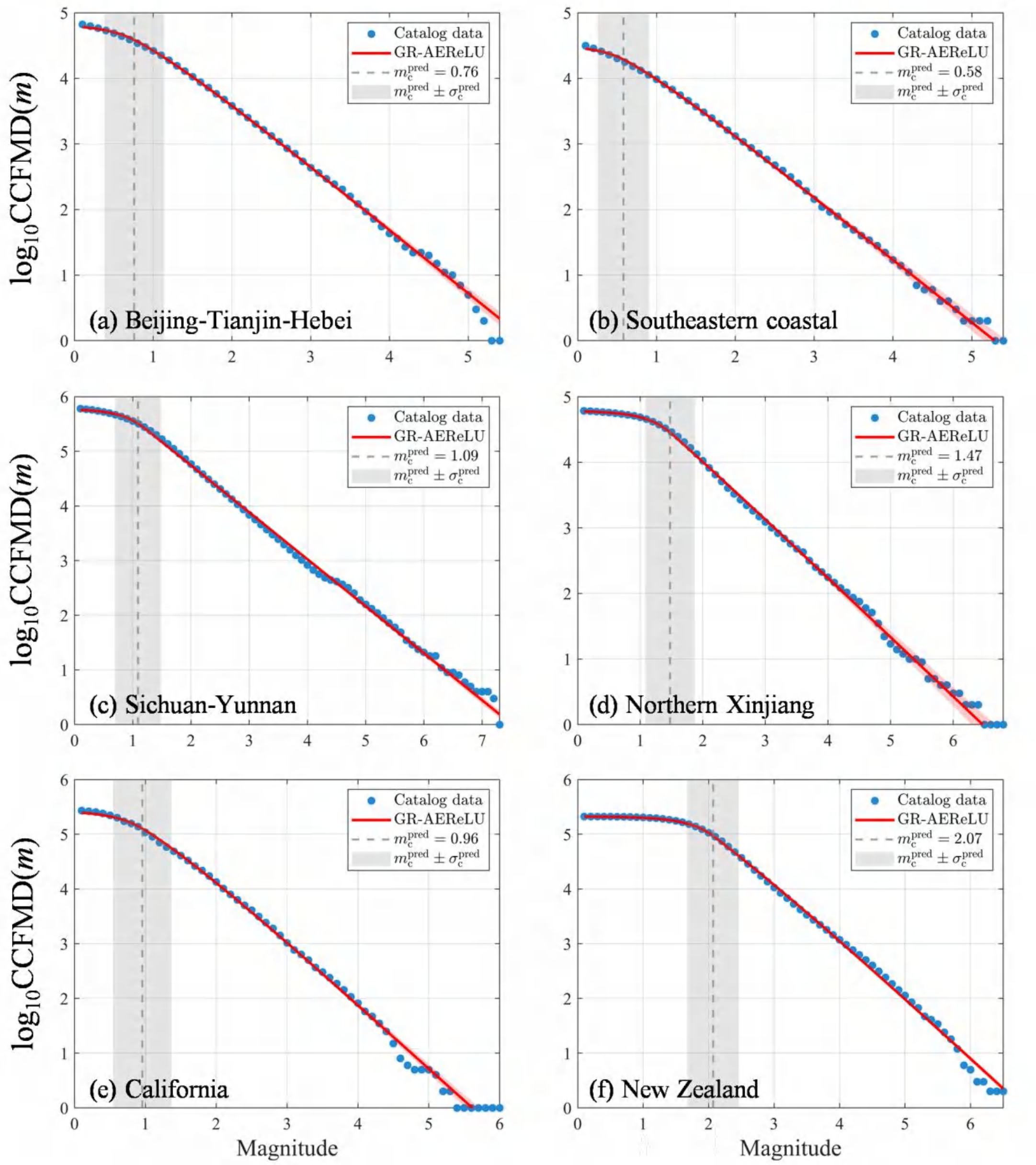

**Figure S10.** CCFMD of the empirical earthquake catalogs from six study regions, along with the fitted CCFMD($m$) derived using the augmented Gutenberg-Richter law with the AEReLU($x$) function. The red shading represents the estimated uncertainty, obtained from 200 bootstrapping iterations.

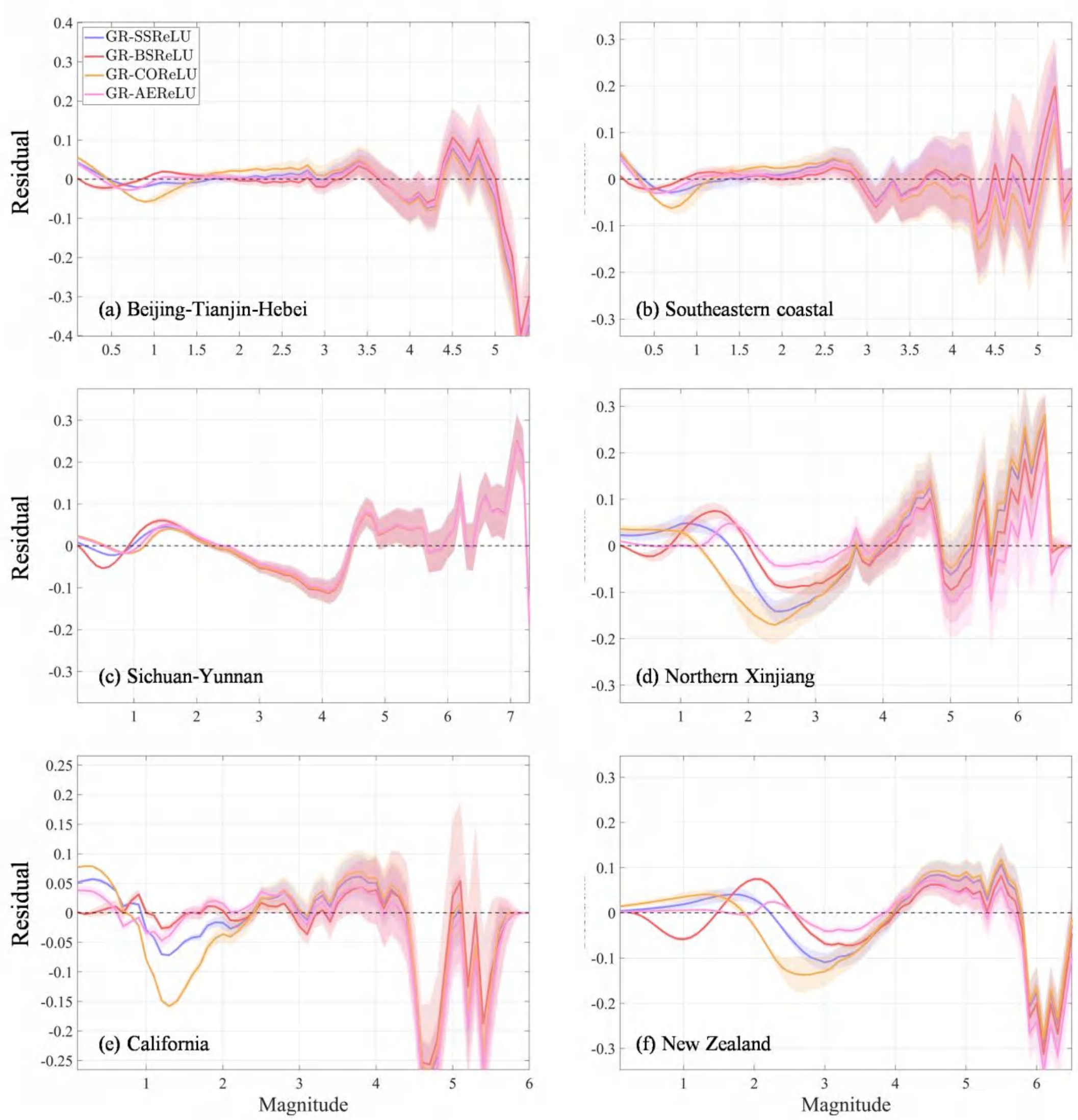

**Figure S11.** Residuals between the empirical $CCFMD^{obs}(m)$ and the fitted augmented Gutenberg-Richter models with four $gReLU(x)$ functions, $CCFMD^{fit}(m)$, expressed as $\log_{10}CCFMD^{obs}(m)$ - $\log_{10}CCFMD^{fit}(m)$, for the six study regions. Shaded areas indicate the uncertainty of the residuals, estimated from 200 bootstrap iterations.

*Support Information (Tables) to*

# Unifying the Gutenberg-Richter Law with Probabilistic Catalog Completeness

Jiawei Li[1], Xinyi Wang[1,2], and Didier Sornette[1]

1 Institute of Risk Analysis, Prediction and Management (Risks-X), Academy for Advanced Interdisciplinary Studies, Southern University of Science and Technology (SUSTech), Shenzhen, China.

2 School of Computer Science, Chengdu University of Information Technology (CUIT), Chengdu, China.

Corresponding authors: Didier Sornette (didier@sustech.edu.cn); Jiawei Li (lijw@cea-igp.ac.cn)

# Jiawei Li and Xinyi Wang contributed equally to this work

**Table S1(a).** Fitted parameters and performance evaluation metrics of the augmented Gutenberg-Richter model with four gReLU($x$) functions applied to simulated catalogs. Each catalog is generated by the GR-SSReLU model using predefined parameters. Among these parameters, the true $a$-value is calculated as $a^{true} = k + 2.5 + 0.75$gReLU(-10/3), based on Equations (1) and (2). Light red shading indicates scenarios with correct model specification; light blue shading indicates scenarios with model mis-specification.

| Model | Predefined parameters | | Fitted parameters | | | | | Rmse | R-Square | SSE | AIC |
|---|---|---|---|---|---|---|---|---|---|---|---|
| | $\theta^{true}$ | $N^{true} = 10^k$ | $a^{pred}$ | $b^{pred}$ | $m_c^{pred}$ | $\sigma_{mc}^{pred}$ | $\beta^{pred}$ | | | | |
| **GR-SSReLU** | $b^{true} = 1.00$, $m_c^{true} = 2.50$, $\sigma_{mc}^{true} = 0.75$ | $k = 3$ | 5.35 ± 0.62 | 0.94 ± 0.15 | 2.45 ± 0.26 | 0.66 ± 0.08 | -- | $(1.25 \pm 0.34)\times10^1$ | 0.9988 ± 0.0007 | $(9.83 \pm 5.51)\times10^3$ | $(3.05 \pm 0.36)\times10^2$ |
| | | $k = 4$ | 6.53 ± 0.42 | 0.98 ± 0.09 | 2.55 ± 0.18 | 0.69 ± 0.06 | -- | $(1.19 \pm 0.33)\times10^2$ | 0.9989 ± 0.0006 | $(1.02 \pm 0.56)\times10^6$ | $(6.58 \pm 0.45)\times10^2$ |
| | | $k = 5$ | 7.60 ± 0.24 | 0.99 ± 0.05 | 2.60 ± 0.11 | 0.71 ± 0.04 | -- | $(1.15 \pm 0.32)\times10^3$ | 0.9989 ± 0.0006 | $(1.07 \pm 0.62)\times10^8$ | $(1.08 \pm 0.04)\times10^3$ |
| | | $k = 6$ | 8.72 ± 0.12 | 1.00 ± 0.02 | 2.66 ± 0.06 | 0.74 ± 0.02 | -- | $(1.31 \pm 0.23)\times10^4$ | 0.9986 ± 0.0005 | $(1.43 \pm 0.51)\times10^{10}$ | $(1.53 \pm 0.03)\times10^3$ |
| **GR-BSReLU** | -- | $k = 3$ | 6.95 ± 1.25 | 1.15 ± 0.20 | 2.84 ± 0.61 | 0.55 ± 0.07 | -- | $(2.74 \pm 0.64)\times10^1$ | 0.9943 ± 0.0027 | $(4.65 \pm 2.05)\times10^4$ | $(4.00 \pm 0.39)\times10^2$ |
| | | $k = 4$ | 7.96 ± 1.04 | 1.17 ± 0.16 | 2.82 ± 0.50 | 0.54 ± 0.07 | -- | $(2.70 \pm 0.49)\times10^2$ | 0.9945 ± 0.0021 | $(5.13 \pm 2.00)\times10^6$ | $(7.74 \pm 0.57)\times10^2$ |
| | | $k = 5$ | 8.86 ± 0.84 | 1.16 ± 0.12 | 2.80 ± 0.40 | 0.54 ± 0.05 | -- | $(2.52 \pm 0.38)\times10^3$ | 0.9951 ± 0.0015 | $(4.98 \pm 1.63)\times10^8$ | $(1.20 \pm 0.06)\times10^3$ |
| | | $k = 6$ | 9.93 ± 0.52 | 1.17 ± 0.07 | 2.85 ± 0.25 | 0.55 ± 0.03 | -- | $(2.31 \pm 0.22)\times10^4$ | 0.9958 ± 0.0008 | $(4.33 \pm 0.84)\times10^{10}$ | $(1.63 \pm 0.03)\times10^3$ |
| **GR-COReLU** | -- | $k = 3$ | 4.78 ± 0.33 | 0.82 ± 0.09 | 2.14 ± 0.15 | 0.95 ± 0.07 | -- | $(2.67 \pm 0.60)\times10^1$ | 0.9946 ± 0.0026 | $(4.42 \pm 2.02)\times10^4$ | $(3.98 \pm 0.36)\times10^2$ |
| | | $k = 4$ | 6.02 ± 0.20 | 0.89 ± 0.05 | 2.27 ± 0.09 | 1.01 ± 0.05 | -- | $(2.98 \pm 0.47)\times10^2$ | 0.9933 ± 0.0020 | $(6.14 \pm 1.68)\times10^6$ | $(7.87 \pm 0.45)\times10^2$ |
| | | $k = 5$ | 7.18 ± 0.11 | 0.93 ± 0.03 | 2.35 ± 0.05 | 1.06 ± 0.03 | -- | $(3.27 \pm 0.31)\times10^3$ | 0.9918 ± 0.0015 | $(8.22 \pm 1.42)\times10^8$ | $(1.24 \pm 0.04)\times10^3$ |
| | | $k = 6$ | 8.32 ± 0.06 | 0.96 ± 0.01 | 2.42 ± 0.03 | 1.11 ± 0.02 | -- | $(3.63 \pm 0.17)\times10^4$ | 0.9898 ± 0.0010 | $(1.06 \pm 0.10)\times10^{11}$ | $(1.70 \pm 0.02)\times10^3$ |
| **GR-AEReLU** | -- | $k = 3$ | 5.32 ± 0.50 | 0.92 ± 0.13 | 2.50 ± 0.23 | 0.86 ± 0.06 | 0.86 ± 0.29 | $(1.94 \pm 0.47)\times10^1$ | 0.9971 ± 0.0015 | $(2.33 \pm 1.11)\times10^4$ | $(3.61 \pm 0.34)\times10^2$ |
| | | $k = 4$ | 6.57 ± 0.43 | 0.98 ± 0.09 | 2.59 ± 0.20 | 0.90 ± 0.05 | 0.89 ± 0.19 | $(2.02 \pm 0.40)\times10^2$ | 0.9969 ± 0.0012 | $(2.86 \pm 1.03)\times10^6$ | $(7.35 \pm 0.42)\times10^2$ |
| | | $k = 5$ | 7.75 ± 0.30 | 1.02 ± 0.05 | 2.69 ± 0.15 | 0.92 ± 0.04 | 0.79 ± 0.12 | $(2.05 \pm 0.32)\times10^3$ | 0.9967 ± 0.0010 | $(3.26 \pm 0.99)\times10^8$ | $(1.17 \pm 0.04)\times10^3$ |
| | | $k = 6$ | 8.98 ± 0.16 | 1.05 ± 0.03 | 2.82 ± 0.08 | 0.95 ± 0.03 | 0.67 ± 0.04 | $(2.20 \pm 0.22)\times10^4$ | 0.9962 ± 0.0008 | $(3.92 \pm 0.79)\times10^{10}$ | $(1.62 \pm 0.02)\times10^3$ |

**Table S1(b).** Performance ranking of the augmented Gutenberg-Richter model with four gReLU($x$) functions applied to simulated catalogs, based on the evaluation metrics shown in Table S1(a).

| **Metrics** | **Predefined parameters** | **Rank** | | | |
|---|---|---|---|---|---|
| | $N^{\text{true}} = 10^k$ | **First** | **Second** | **Third** | **Fourth** |
| **Rmse** | $k = 3$ | GR-SSReLU | GR-AEReLU | GR-COReLU | GR-BSReLU |
| | $k = 4$ | GR-SSReLU | GR-AEReLU | GR-BSReLU | GR-COReLU |
| | $k = 5$ | GR-SSReLU | GR-AEReLU | GR-BSReLU | GR-COReLU |
| | $k = 6$ | GR-SSReLU | GR-AEReLU | GR-BSReLU | GR-COReLU |
| **R-Square** | $k = 3$ | GR-SSReLU | GR-AEReLU | GR-COReLU | GR-BSReLU |
| | $k = 4$ | GR-SSReLU | GR-AEReLU | GR-BSReLU | GR-COReLU |
| | $k = 5$ | GR-SSReLU | GR-AEReLU | GR-BSReLU | GR-COReLU |
| | $k = 6$ | GR-SSReLU | GR-AEReLU | GR-BSReLU | GR-COReLU |
| **SSE** | $k = 3$ | GR-SSReLU | GR-AEReLU | GR-COReLU | GR-BSReLU |
| | $k = 4$ | GR-SSReLU | GR-AEReLU | GR-BSReLU | GR-COReLU |
| | $k = 5$ | GR-SSReLU | GR-AEReLU | GR-BSReLU | GR-COReLU |
| | $k = 6$ | GR-SSReLU | GR-AEReLU | GR-BSReLU | GR-COReLU |
| **AIC** | $k = 3$ | GR-SSReLU | GR-AEReLU | GR-COReLU | GR-BSReLU |
| | $k = 4$ | GR-SSReLU | GR-AEReLU | GR-BSReLU | GR-COReLU |
| | $k = 5$ | GR-SSReLU | GR-AEReLU | GR-BSReLU | GR-COReLU |
| | $k = 6$ | GR-SSReLU | GR-AEReLU | GR-BSReLU | GR-COReLU |

**Table S2(a).** Fitted parameters and performance evaluation metrics of the augmented Gutenberg-Richter model with four gReLU($x$) functions applied to simulated catalogs. Each catalog is generated by the GR-BSReLU model using predefined parameters. The details are same as Table S1(a).

| **Model** | **Predefined parameters** | | **Fitted parameters** | | | | | **Rmse** | **R-Square** | **SSE** | **AIC** |
|---|---|---|---|---|---|---|---|---|---|---|---|
| | $\theta^{true}$ | $N^{true} = 10^k$ | $a^{pred}$ | $b^{pred}$ | $m_c^{pred}$ | $\sigma_{mc}^{pred}$ | $\beta^{pred}$ | | | | |
| **GR-SSReLU** | -- | $k = 3$ | 4.40 ± 0.28 | 0.80 ± 0.09 | 1.74 ± 0.15 | 0.44 ± 0.06 | -- | $(1.91 \pm 0.60)\times10^1$ | 0.9970 ± 0.0018 | $(2.10 \pm 1.20)\times10^4$ | $(3.20 \pm 0.37)\times10^2$ |
| | | $k = 4$ | 5.58 ± 0.20 | 0.85 ± 0.05 | 1.86 ± 0.11 | 0.51 ± 0.05 | -- | $(2.36 \pm 0.61)\times10^2$ | 0.9955 ± 0.0021 | $(3.78 \pm 1.75)\times10^6$ | $(7.13 \pm 0.46)\times10^2$ |
| | | $k = 5$ | 6.71 ± 0.14 | 0.88 ± 0.03 | 1.95 ± 0.08 | 0.56 ± 0.04 | -- | $(2.75 \pm 0.50)\times10^3$ | 0.9939 ± 0.0021 | $(5.75 \pm 1.94)\times10^8$ | $(1.18 \pm 0.05)\times10^3$ |
| | | $k = 6$ | 7.90 ± 0.08 | 0.92 ± 0.02 | 2.06 ± 0.05 | 0.63 ± 0.03 | -- | $(3.46 \pm 0.31)\times10^4$ | 0.9907 ± 0.0017 | $(9.66 \pm 1.72)\times10^{10}$ | $(1.68 \pm 0.03)\times10^3$ |
| **GR-BSReLU** | $b^{true} = 1.00$<br>$m_c^{true} = 2.50$<br>$\sigma_{mc}^{true} = 0.75$ | $k = 3$ | 5.27 ± 1.18 | 0.94 ± 0.21 | 1.98 ± 0.79 | 0.53 ± 0.16 | -- | $(8.00 \pm 3.18)\times10^0$ | 0.9994 ± 0.0005 | $(3.99 \pm 3.51)\times10^3$ | $(2.55 \pm 0.47)\times10^2$ |
| | | $k = 4$ | 6.59 ± 1.19 | 0.99 ± 0.17 | 2.27 ± 0.87 | 0.60 ± 0.17 | -- | $(5.13 \pm 2.79)\times10^1$ | 0.9997 ± 0.0003 | $(2.22 \pm 2.44)\times10^5$ | $(5.00 \pm 0.85)\times10^2$ |
| | | $k = 5$ | 7.54 ± 0.84 | 0.99 ± 0.11 | 2.33 ± 0.70 | 0.63 ± 0.15 | -- | $(4.15 \pm 2.41)\times10^2$ | 0.9998 ± 0.0002 | $(1.69 \pm 1.97)\times10^7$ | $(8.78 \pm 1.00)\times10^2$ |
| | | $k = 6$ | 9.13 ± 0.65 | 1.06 ± 0.07 | 2.91 ± 0.42 | 0.77 ± 0.07 | -- | $(5.62 \pm 2.92)\times10^3$ | 0.9997 ± 0.0003 | $(3.20 \pm 3.03)\times10^9$ | $(1.36 \pm 0.10)\times10^3$ |
| **GR-COReLU** | -- | $k = 3$ | 4.22 ± 0.15 | 0.76 ± 0.06 | 1.61 ± 0.09 | 0.67 ± 0.06 | -- | $(3.31 \pm 0.61)\times10^1$ | 0.9914 ± 0.0031 | $(5.99 \pm 2.05)\times10^4$ | $(3.83 \pm 0.35)\times10^2$ |
| | | $k = 4$ | 5.38 ± 0.11 | 0.80 ± 0.03 | 1.71 ± 0.07 | 0.75 ± 0.05 | -- | $(3.78 \pm 0.52)\times10^2$ | 0.9889 ± 0.0028 | $(9.31 \pm 2.23)\times10^6$ | $(7.77 \pm 0.50)\times10^2$ |
| | | $k = 5$ | 6.51 ± 0.08 | 0.84 ± 0.02 | 1.80 ± 0.05 | 0.83 ± 0.04 | -- | $(4.26 \pm 0.39)\times10^3$ | 0.9856 ± 0.0025 | $(1.35 \pm 0.22)\times10^9$ | $(1.25 \pm 0.05)\times10^3$ |
| | | $k = 6$ | 7.66 ± 0.05 | 0.88 ± 0.01 | 1.89 ± 0.03 | 0.92 ± 0.02 | -- | $(4.94 \pm 0.22)\times10^4$ | 0.9812 ± 0.0017 | $(1.96 \pm 0.17)\times10^{11}$ | $(1.74 \pm 0.02)\times10^3$ |
| **GR-AEReLU** | -- | $k = 3$ | 4.48 ± 0.26 | 0.81 ± 0.08 | 1.82 ± 0.15 | 0.56 ± 0.07 | 0.67 ± 0.22 | $(2.21 \pm 0.58)\times10^1$ | 0.9960 ± 0.0020 | $(2.76 \pm 1.34)\times10^4$ | $(3.40 \pm 0.34)\times10^2$ |
| | | $k = 4$ | 5.70 ± 0.22 | 0.87 ± 0.05 | 1.95 ± 0.13 | 0.63 ± 0.06 | 0.65 ± 0.09 | $(2.57 \pm 0.57)\times10^2$ | 0.9947 ± 0.0022 | $(4.42 \pm 1.76)\times10^6$ | $(7.27 \pm 0.46)\times10^2$ |
| | | $k = 5$ | 6.87 ± 0.18 | 0.90 ± 0.04 | 2.07 ± 0.11 | 0.68 ± 0.05 | 0.59 ± 0.05 | $(2.87 \pm 0.48)\times10^3$ | 0.9933 ± 0.0021 | $(6.23 \pm 1.94)\times10^8$ | $(1.19 \pm 0.05)\times10^3$ |
| | | $k = 6$ | 8.13 ± 0.10 | 0.95 ± 0.02 | 2.24 ± 0.06 | 0.76 ± 0.03 | 0.54 ± 0.01 | $(3.52 \pm 0.30)\times10^4$ | 0.9904 ± 0.0016 | $(9.95 \pm 1.68)\times10^{10}$ | $(1.68 \pm 0.03)\times10^3$ |

| **Table S2(b).** Performance ranking of the augmented Gutenberg-Richter model with four gReLU($x$) functions applied to simulated catalogs, based on the evaluation metrics shown in Table S2(a). | | | | | |
|---|---|---|---|---|---|
| **Metrics** | **Predefined parameters** | **Rank** | | | |
| | $N^{true} = 10^k$ | **First** | **Second** | **Third** | **Fourth** |
| **Rmse** | $k = 3$ | GR-BSReLU | GR-SSReLU | GR-AEReLU | GR-COReLU |
| | $k = 4$ | GR-BSReLU | GR-SSReLU | GR-AEReLU | GR-COReLU |
| | $k = 5$ | GR-BSReLU | GR-SSReLU | GR-AEReLU | GR-COReLU |
| | $k = 6$ | GR-BSReLU | GR-SSReLU | GR-AEReLU | GR-COReLU |
| **R-Square** | $k = 3$ | GR-BSReLU | GR-SSReLU | GR-AEReLU | GR-COReLU |
| | $k = 4$ | GR-BSReLU | GR-SSReLU | GR-AEReLU | GR-COReLU |
| | $k = 5$ | GR-BSReLU | GR-SSReLU | GR-AEReLU | GR-COReLU |
| | $k = 6$ | GR-BSReLU | GR-SSReLU | GR-AEReLU | GR-COReLU |
| **SSE** | $k = 3$ | GR-BSReLU | GR-SSReLU | GR-AEReLU | GR-COReLU |
| | $k = 4$ | GR-BSReLU | GR-SSReLU | GR-AEReLU | GR-COReLU |
| | $k = 5$ | GR-BSReLU | GR-SSReLU | GR-AEReLU | GR-COReLU |
| | $k = 6$ | GR-BSReLU | GR-SSReLU | GR-AEReLU | GR-COReLU |
| **AIC** | $k = 3$ | GR-BSReLU | GR-SSReLU | GR-AEReLU | GR-COReLU |
| | $k = 4$ | GR-BSReLU | GR-SSReLU | GR-AEReLU | GR-COReLU |
| | $k = 5$ | GR-BSReLU | GR-SSReLU | GR-AEReLU | GR-COReLU |
| | $k = 6$ | GR-BSReLU | GR-SSReLU | GR-AEReLU | GR-COReLU |

**Table S3(a).** Fitted parameters and performance evaluation metrics of the augmented Gutenberg-Richter model with four gReLU($x$) functions applied to simulated catalogs. Each catalog is generated by the GR-COReLU model using predefined parameters. The details are same as Table S1(a).

| **Model** | **Predefined parameters** | | **Fitted parameters** | | | | | **Rmse** | **R-Square** | **SSE** | **AIC** |
|---|---|---|---|---|---|---|---|---|---|---|---|
| | $\theta^{true}$ | $N^{true} = 10^k$ | $a^{pred}$ | $b^{pred}$ | $m_c^{pred}$ | $\sigma_{mc}^{pred}$ | $\beta^{pred}$ | | | | |
| **GR-SSReLU** | -- | $k = 3$ | 6.02 ± 0.53 | 1.08 ± 0.13 | 2.78 ± 0.16 | 0.48 ± 0.06 | -- | $(1.86 \pm 0.66)\times10^1$ | 0.9976 ± 0.0016 | $(2.33 \pm 1.62)\times10^4$ | $(3.49 \pm 0.62)\times10^2$ |
| | | $k = 4$ | 6.88 ± 0.32 | 1.05 ± 0.08 | 2.73 ± 0.11 | 0.46 ± 0.04 | -- | $(2.02 \pm 0.51)\times10^2$ | 0.9973 ± 0.0013 | $(3.01 \pm 1.56)\times10^6$ | $(7.32 \pm 0.73)\times10^2$ |
| | | $k = 5$ | 7.79 ± 0.17 | 1.03 ± 0.04 | 2.70 ± 0.06 | 0.44 ± 0.03 | -- | $(2.05 \pm 0.38)\times10^3$ | 0.9973 ± 0.0010 | $(3.32 \pm 1.25)\times10^8$ | $(1.17 \pm 0.06)\times10^3$ |
| | | $k = 6$ | 8.81 ± 0.09 | 1.04 ± 0.02 | 2.71 ± 0.04 | 0.45 ± 0.02 | -- | $(1.93 \pm 0.26)\times10^4$ | 0.9976 ± 0.0006 | $(3.06 \pm 0.80)\times10^{10}$ | $(1.60 \pm 0.03)\times10^3$ |
| **GR-BSReLU** | -- | $k = 3$ | 6.50 ± 0.59 | 1.16 ± 0.13 | 2.61 ± 0.28 | 0.33 ± 0.06 | -- | $(3.61 \pm 0.55)\times10^1$ | 0.9917 ± 0.0028 | $(7.81 \pm 2.29)\times10^4$ | $(4.32 \pm 0.38)\times10^2$ |
| | | $k = 4$ | 7.30 ± 0.37 | 1.13 ± 0.08 | 2.59 ± 0.26 | 0.32 ± 0.05 | -- | $(3.45 \pm 0.27)\times10^2$ | 0.9927 ± 0.0012 | $(8.13 \pm 1.22)\times10^6$ | $(8.08 \pm 0.55)\times10^2$ |
| | | $k = 5$ | 8.17 ± 0.18 | 1.10 ± 0.04 | 2.68 ± 0.27 | 0.34 ± 0.05 | -- | $(3.26 \pm 0.17)\times10^3$ | 0.9934 ± 0.0007 | $(8.09 \pm 0.78)\times10^8$ | $(1.24 \pm 0.05)\times10^3$ |
| | | $k = 6$ | 9.15 ± 0.08 | 1.09 ± 0.02 | 2.79 ± 0.21 | 0.35 ± 0.03 | -- | $(3.16 \pm 0.09)\times10^4$ | 0.9937 ± 0.0004 | $(8.05 \pm 0.46)\times10^{10}$ | $(1.68 \pm 0.02)\times10^3$ |
| **GR-COReLU** | $b^{true} = 1.00$ $m_c^{true} = 2.50$ $\sigma_{mc}^{true} = 0.75$ | $k = 3$ | 5.71 ± 0.41 | 1.02 ± 0.11 | 2.64 ± 0.13 | 0.76 ± 0.07 | -- | $(1.28 \pm 0.81)\times10^1$ | 0.9986 ± 0.0024 | $(1.37 \pm 2.33)\times10^4$ | $(2.95 \pm 0.66)\times10^2$ |
| | | $k = 4$ | 6.68 ± 0.20 | 1.02 ± 0.05 | 2.64 ± 0.07 | 0.74 ± 0.04 | -- | $(8.69 \pm 4.46)\times10^1$ | 0.9994 ± 0.0007 | $(6.50 \pm 8.26)\times10^5$ | $(6.06 \pm 0.70)\times10^2$ |
| | | $k = 5$ | 7.66 ± 0.11 | 1.01 ± 0.03 | 2.63 ± 0.04 | 0.74 ± 0.03 | -- | $(7.48 \pm 3.53)\times10^2$ | 0.9996 ± 0.0005 | $(5.19 \pm 5.66)\times10^7$ | $(1.00 \pm 0.07)\times10^3$ |
| | | $k = 6$ | 8.67 ± 0.06 | 1.01 ± 0.01 | 2.63 ± 0.02 | 0.74 ± 0.01 | -- | $(6.25 \pm 1.46)\times10^3$ | 0.9997 ± 0.0002 | $(3.32 \pm 1.95)\times10^9$ | $(1.41 \pm 0.03)\times10^3$ |
| **GR-AEReLU** | -- | $k = 3$ | 6.05 ± 0.61 | 1.05 ± 0.15 | 2.90 ± 0.29 | 0.72 ± 0.04 | 1.05 ± 0.85 | $(9.06 \pm 3.01)\times10^0$ | 0.9994 ± 0.0004 | $(5.50 \pm 4.30)\times10^3$ | $(2.68 \pm 0.50)\times10^2$ |
| | | $k = 4$ | 6.86 ± 0.44 | 1.05 ± 0.09 | 2.72 ± 0.18 | 0.74 ± 0.04 | 1.59 ± 0.57 | $(7.94 \pm 2.51)\times10^1$ | 0.9996 ± 0.0003 | $(4.80 \pm 3.34)\times10^5$ | $(6.04 \pm 0.69)\times10^2$ |
| | | $k = 5$ | 7.76 ± 0.28 | 1.03 ± 0.05 | 2.68 ± 0.12 | 0.74 ± 0.04 | 1.67 ± 0.43 | $(7.57 \pm 2.09)\times10^2$ | 0.9996 ± 0.0002 | $(4.72 \pm 2.84)\times10^7$ | $(1.02 \pm 0.07)\times10^3$ |
| | | $k = 6$ | 8.79 ± 0.23 | 1.03 ± 0.03 | 2.70 ± 0.12 | 0.73 ± 0.03 | 1.58 ± 0.37 | $(6.48 \pm 1.32)\times10^3$ | 0.9997 ± 0.0001 | $(3.52 \pm 1.48)\times10^9$ | $(1.42 \pm 0.04)\times10^3$ |

**Table S3(b).** Performance ranking of the augmented Gutenberg-Richter model with four gReLU($x$) functions applied to simulated catalogs, based on the evaluation metrics shown in Table S3(a).

| **Metrics** | **Predefined parameters** | **Rank** | | | |
|---|---|---|---|---|---|
| | $N^{true} = 10^k$ | **First** | **Second** | **Third** | **Fourth** |
| **Rmse** | $k = 3$ | GR-AEReLU | GR-COReLU | GR-SSReLU | GR-BSReLU |
| | $k = 4$ | GR-AEReLU | GR-COReLU | GR-SSReLU | GR-BSReLU |
| | $k = 5$ | GR-COReLU | GR-AEReLU | GR-SSReLU | GR-BSReLU |
| | $k = 6$ | GR-COReLU | GR-AEReLU | GR-SSReLU | GR-BSReLU |
| **R-Square** | $k = 3$ | GR-AEReLU | GR-COReLU | GR-SSReLU | GR-BSReLU |
| | $k = 4$ | GR-AEReLU | GR-COReLU | GR-SSReLU | GR-BSReLU |
| | $k = 5$ | GR-AEReLU | GR-COReLU | GR-SSReLU | GR-BSReLU |
| | $k = 6$ | GR-COReLU | GR-AEReLU | GR-SSReLU | GR-BSReLU |
| **SSE** | $k = 3$ | GR-AEReLU | GR-COReLU | GR-SSReLU | GR-BSReLU |
| | $k = 4$ | GR-AEReLU | GR-COReLU | GR-SSReLU | GR-BSReLU |
| | $k = 5$ | GR-AEReLU | GR-COReLU | GR-SSReLU | GR-BSReLU |
| | $k = 6$ | GR-COReLU | GR-AEReLU | GR-SSReLU | GR-BSReLU |
| **AIC** | $k = 3$ | GR-AEReLU | GR-COReLU | GR-SSReLU | GR-BSReLU |
| | $k = 4$ | GR-AEReLU | GR-COReLU | GR-SSReLU | GR-BSReLU |
| | $k = 5$ | GR-COReLU | GR-AEReLU | GR-SSReLU | GR-BSReLU |
| | $k = 6$ | GR-COReLU | GR-AEReLU | GR-SSReLU | GR-BSReLU |

**Table S4(a).** Fitted parameters and performance evaluation metrics of the augmented Gutenberg-Richter model with four gReLU($x$) functions applied to simulated catalogs. Each catalog is generated by the GR-AEReLU model using predefined parameters. The details are same as Table S1(a).

| Model | Predefined parameters | | Fitted parameters | | | | | Rmse | R-Square | SSE | AIC |
|---|---|---|---|---|---|---|---|---|---|---|---|
| | $\theta^{true}$ | $N^{true} = 10^k$ | $a^{pred}$ | $b^{pred}$ | $m_c^{pred}$ | $\sigma_{mc}^{pred}$ | $\beta^{pred}$ | | | | |
| **GR-SSReLU** | -- | $k = 3$ | 4.98 ± 0.39 | 0.91 ± 0.11 | 2.17 ± 0.16 | 0.50 ± 0.06 | -- | (1.27 ± 0.53)×10$^1$ | 0.9987 ± 0.0012 | (1.12 ± 1.08)×10$^4$ | (2.99 ± 0.62)×10$^2$ |
| | | $k = 4$ | 5.97 ± 0.20 | 0.91 ± 0.05 | 2.17 ± 0.09 | 0.49 ± 0.04 | -- | (9.96 ± 3.23)×10$^1$ | 0.9992 ± 0.0005 | (7.53 ± 5.67)×10$^5$ | (6.27 ± 0.69)×10$^2$ |
| | | $k = 5$ | 7.03 ± 0.13 | 0.92 ± 0.03 | 2.20 ± 0.07 | 0.51 ± 0.03 | -- | (8.02 ± 1.82)×10$^2$ | 0.9995 ± 0.0003 | (5.15 ± 2.83)×10$^7$ | (1.02 ± 0.05)×10$^3$ |
| | | $k = 6$ | 8.15 ± 0.08 | 0.95 ± 0.02 | 2.26 ± 0.04 | 0.55 ± 0.02 | -- | (7.82 ± 1.61)×10$^3$ | 0.9995 ± 0.0002 | (5.10 ± 2.30)×10$^9$ | (1.44 ± 0.03)×10$^3$ |
| **GR-BSReLU** | -- | $k = 3$ | 5.58 ± 0.70 | 1.01 ± 0.16 | 2.14 ± 0.32 | 0.44 ± 0.08 | -- | (2.80 ± 0.45)×10$^1$ | 0.9946 ± 0.0017 | (4.69 ± 1.52)×10$^4$ | (3.99 ± 0.44)×10$^2$ |
| | | $k = 4$ | 6.40 ± 0.43 | 0.98 ± 0.09 | 2.09 ± 0.28 | 0.43 ± 0.07 | -- | (2.59 ± 0.25)×10$^2$ | 0.9954 ± 0.0009 | (4.59 ± 0.94)×10$^6$ | (7.61 ± 0.54)×10$^2$ |
| | | $k = 5$ | 7.39 ± 0.33 | 0.98 ± 0.06 | 2.10 ± 0.31 | 0.43 ± 0.06 | -- | (2.43 ± 0.16)×10$^3$ | 0.9958 ± 0.0005 | (4.51 ± 0.62)×10$^8$ | (1.19 ± 0.05)×10$^3$ |
| | | $k = 6$ | 8.50 ± 0.25 | 1.00 ± 0.04 | 2.20 ± 0.34 | 0.46 ± 0.07 | -- | (2.31 ± 0.10)×10$^4$ | 0.9961 ± 0.0003 | (4.28 ± 0.36)×10$^{10}$ | (1.62 ± 0.02)×10$^3$ |
| **GR-COReLU** | -- | $k = 3$ | 4.79 ± 0.26 | 0.87 ± 0.08 | 2.06 ± 0.12 | 0.78 ± 0.08 | -- | (1.99 ± 0.98)×10$^1$ | 0.9967 ± 0.0035 | (2.91 ± 3.13)×10$^4$ | (3.48 ± 0.68)×10$^2$ |
| | | $k = 4$ | 5.83 ± 0.14 | 0.88 ± 0.04 | 2.08 ± 0.07 | 0.79 ± 0.05 | -- | (1.82 ± 0.72)×10$^2$ | 0.9974 ± 0.0021 | (2.59 ± 2.14)×10$^6$ | (7.04 ± 0.66)×10$^2$ |
| | | $k = 5$ | 6.88 ± 0.08 | 0.90 ± 0.02 | 2.10 ± 0.04 | 0.80 ± 0.03 | -- | (1.79 ± 0.41)×10$^3$ | 0.9976 ± 0.0012 | (2.56 ± 1.30)×10$^8$ | (1.14 ± 0.05)×10$^3$ |
| | | $k = 6$ | 7.95 ± 0.05 | 0.91 ± 0.01 | 2.13 ± 0.02 | 0.83 ± 0.02 | -- | (1.98 ± 0.19)×10$^4$ | 0.9971 ± 0.0005 | (3.17 ± 0.59)×10$^{10}$ | (1.60 ± 0.02)×10$^3$ |
| **GR-AEReLU** | $b^{true} = 1.00$ | $k = 3$ | 5.12 ± 0.39 | 0.89 ± 0.12 | 2.36 ± 0.22 | 0.69 ± 0.03 | 0.68 ± 0.62 | (1.01 ± 0.23)×10$^1$ | 0.9993 ± 0.0003 | (6.27 ± 2.88)×10$^3$ | (2.80 ± 0.36)×10$^2$ |
| | $m_c^{true} = 2.50$ | $k = 4$ | 6.12 ± 0.33 | 0.92 ± 0.08 | 2.28 ± 0.21 | 0.68 ± 0.03 | 0.88 ± 0.47 | (7.91 ± 1.39)×10$^1$ | 0.9996 ± 0.0002 | (4.35 ± 1.64)×10$^5$ | (6.01 ± 0.43)×10$^2$ |
| | $\sigma_{mc}^{true} = 0.75$ | $k = 5$ | 7.27 ± 0.30 | 0.96 ± 0.05 | 2.36 ± 0.18 | 0.68 ± 0.03 | 0.70 ± 0.31 | (7.28 ± 1.57)×10$^2$ | 0.9996 ± 0.0002 | (4.17 ± 2.11)×10$^7$ | (1.01 ± 0.04)×10$^3$ |
| | $\beta^{true} = 0.35$ | $k = 6$ | 8.55 ± 0.13 | 1.00 ± 0.02 | 2.54 ± 0.08 | 0.71 ± 0.03 | 0.44 ± 0.10 | (8.64 ± 1.78)×10$^3$ | 0.9994 ± 0.0002 | (6.23 ± 2.68)×10$^9$ | (1.46 ± 0.03)×10$^3$ |

**Table S4(b).** Performance ranking of the augmented Gutenberg-Richter model with four gReLU($x$) functions applied to simulated catalogs, based on the evaluation metrics shown in Table S4(a).

| **Metrics** | **Predefined parameters** | **Rank** | | | |
|---|---|---|---|---|---|
| | $N^{true} = 10^k$ | **First** | **Second** | **Third** | **Fourth** |
| **Rmse** | $k = 3$ | GR-AEReLU | GR-SSReLU | GR-COReLU | GR-BSReLU |
| | $k = 4$ | GR-AEReLU | GR-SSReLU | GR-COReLU | GR-BSReLU |
| | $k = 5$ | GR-AEReLU | GR-SSReLU | GR-COReLU | GR-BSReLU |
| | $k = 6$ | GR-SSReLU | GR-AEReLU | GR-COReLU | GR-BSReLU |
| **R-Square** | $k = 3$ | GR-AEReLU | GR-SSReLU | GR-COReLU | GR-BSReLU |
| | $k = 4$ | GR-AEReLU | GR-SSReLU | GR-COReLU | GR-BSReLU |
| | $k = 5$ | GR-AEReLU | GR-SSReLU | GR-COReLU | GR-BSReLU |
| | $k = 6$ | GR-SSReLU | GR-AEReLU | GR-COReLU | GR-BSReLU |
| **SSE** | $k = 3$ | GR-AEReLU | GR-SSReLU | GR-COReLU | GR-BSReLU |
| | $k = 4$ | GR-AEReLU | GR-SSReLU | GR-COReLU | GR-BSReLU |
| | $k = 5$ | GR-AEReLU | GR-SSReLU | GR-COReLU | GR-BSReLU |
| | $k = 6$ | GR-SSReLU | GR-AEReLU | GR-COReLU | GR-BSReLU |
| **AIC** | $k = 3$ | GR-AEReLU | GR-SSReLU | GR-COReLU | GR-BSReLU |
| | $k = 4$ | GR-AEReLU | GR-SSReLU | GR-COReLU | GR-BSReLU |
| | $k = 5$ | GR-AEReLU | GR-SSReLU | GR-COReLU | GR-BSReLU |
| | $k = 6$ | GR-SSReLU | GR-AEReLU | GR-COReLU | GR-BSReLU |

**Table S5(a).** Performance evaluation metrics of the augmented Gutenberg-Richter model with four gReLU($x$) functions applied to empirical earthquake catalogs from six study regions.

| Model | Region | Rmse | R-Square | SSE | AIC | $p$-value |
|---|---|---|---|---|---|---|
| **GR-SSReLU** | Beijing-Tianjin-Hebei | $(9.06 \pm 1.62)\times10^{2}$ | $0.9989 \pm 0.0004$ | $(1.30 \pm 0.47)\times10^{8}$ | $(2.10 \pm 0.05)\times10^{3}$ | 0.2814 |
| | Southeastern coastal | $(4.92 \pm 1.09)\times10^{2}$ | $0.9986 \pm 0.0006$ | $(3.88 \pm 1.71)\times10^{7}$ | $(1.91 \pm 0.06)\times10^{3}$ | 1.0000 |
| | Sichuan-Yunnan | $(5.54 \pm 0.79)\times10^{3}$ | $0.9996 \pm 0.0001$ | $(5.44 \pm 1.51)\times10^{9}$ | $(3.00 \pm 0.06)\times10^{3}$ | 0.8650 |
| | Northern Xinjiang | $(6.55 \pm 1.70)\times10^{2}$ | $0.9994 \pm 0.0003$ | $(7.65 \pm 3.89)\times10^{7}$ | $(2.16 \pm 0.11)\times10^{3}$ | 1.0000 |
| | California | $(5.01 \pm 1.12)\times10^{3}$ | $0.9981 \pm 0.0008$ | $(4.14 \pm 1.64)\times10^{9}$ | $(2.69 \pm 0.08)\times10^{3}$ | 1.0000 |
| | New Zealand | $(1.03 \pm 0.30)\times10^{3}$ | $0.9999 \pm 0.0001$ | $(1.90 \pm 1.14)\times10^{8}$ | $(2.29 \pm 0.10)\times10^{3}$ | 0.3859 |
| **GR-BSReLU** | Beijing-Tianjin-Hebei | $(2.64 \pm 0.45)\times10^{2}$ | $0.9999 \pm 0.0000$ | $(1.10 \pm 0.40)\times10^{7}$ | $(1.72 \pm 0.05)\times10^{3}$ | 0.3461 |
| | Southeastern coastal | $(2.07 \pm 0.36)\times10^{2}$ | $0.9998 \pm 0.0001$ | $(6.79 \pm 2.45)\times10^{6}$ | $(1.64 \pm 0.05)\times10^{3}$ | 0.9447 |
| | Sichuan-Yunnan | $(1.15 \pm 0.24)\times10^{4}$ | $0.9982 \pm 0.0008$ | $(2.39 \pm 1.02)\times10^{10}$ | $(3.25 \pm 0.08)\times10^{3}$ | 0.5274 |
| | Northern Xinjiang | $(1.14 \pm 0.34)\times10^{3}$ | $0.9980 \pm 0.0012$ | $(2.38 \pm 1.53)\times10^{8}$ | $(2.35 \pm 0.11)\times10^{3}$ | 1.0000 |
| | California | $(1.36 \pm 0.35)\times10^{3}$ | $0.9999 \pm 0.0001$ | $(3.09 \pm 1.63)\times10^{8}$ | $(2.28 \pm 0.07)\times10^{3}$ | 1.0000 |
| | New Zealand | $(2.75 \pm 0.71)\times10^{3}$ | $0.9990 \pm 0.0005$ | $(1.34 \pm 0.69)\times10^{9}$ | $(2.62 \pm 0.09)\times10^{3}$ | 0.5813 |
| **GR-COReLU** | Beijing-Tianjin-Hebei | $(1.46 \pm 0.13)\times10^{3}$ | $0.9973 \pm 0.0005$ | $(3.29 \pm 0.57)\times10^{8}$ | $(2.25 \pm 0.03)\times10^{3}$ | 0.2408 |
| | Southeastern coastal | $(6.80 \pm 1.07)\times10^{2}$ | $0.9974 \pm 0.0008$ | $(7.26 \pm 2.14)\times10^{7}$ | $(2.01 \pm 0.04)\times10^{3}$ | 1.0000 |
| | Sichuan-Yunnan | $(5.99 \pm 0.52)\times10^{3}$ | $0.9995 \pm 0.0001$ | $(6.28 \pm 1.17)\times10^{9}$ | $(3.03 \pm 0.03)\times10^{3}$ | 0.8738 |
| | Northern Xinjiang | $(4.07 \pm 1.44)\times10^{2}$ | $0.9997 \pm 0.0002$ | $(3.12 \pm 2.92)\times10^{7}$ | $(2.00 \pm 0.11)\times10^{3}$ | 1.0000 |
| | California | $(6.31 \pm 0.85)\times10^{3}$ | $0.9971 \pm 0.0008$ | $(6.39 \pm 1.60)\times10^{9}$ | $(2.77 \pm 0.06)\times10^{3}$ | 0.7772 |
| | New Zealand | $(1.18 \pm 0.20)\times10^{3}$ | $0.9998 \pm 0.0001$ | $(2.36 \pm 0.79)\times10^{8}$ | $(2.35 \pm 0.06)\times10^{3}$ | 0.3580 |
| **GR-AEReLU** | Beijing-Tianjin-Hebei | $(9.09 \pm 1.59)\times10^{2}$ | $0.9989 \pm 0.0004$ | $(1.31 \pm 0.49)\times10^{8}$ | $(2.11 \pm 0.05)\times10^{3}$ | 0.3050 |
| | Southeastern coastal | $(4.59 \pm 1.13)\times10^{2}$ | $0.9988 \pm 0.0006$ | $(3.43 \pm 1.72)\times10^{7}$ | $(1.89 \pm 0.07)\times10^{3}$ | 0.9961 |
| | Sichuan-Yunnan | $(5.79 \pm 0.34)\times10^{3}$ | $0.9996 \pm 0.0001$ | $(5.85 \pm 0.72)\times10^{9}$ | $(3.02 \pm 0.03)\times10^{3}$ | 0.8942 |
| | Northern Xinjiang | $(3.96 \pm 1.16)\times10^{2}$ | $0.9998 \pm 0.0002$ | $(2.85 \pm 2.14)\times10^{7}$ | $(1.99 \pm 0.10)\times10^{3}$ | 1.0000 |
| | California | $(4.50 \pm 1.04)\times10^{3}$ | $0.9985 \pm 0.0007$ | $(3.36 \pm 1.39)\times10^{9}$ | $(2.66 \pm 0.08)\times10^{3}$ | 1.0000 |
| | New Zealand | $(9.46 \pm 1.14)\times10^{2}$ | $0.9999 \pm 0.0000$ | $(1.50 \pm 0.37)\times10^{8}$ | $(2.28 \pm 0.04)\times10^{3}$ | 0.5240 |

**Table S5(b).** Performance ranking of the augmented Gutenberg-Richter model with four gReLU(*x*) functions applied to empirical catalogs, based on the evaluation metrics shown in Table S5(a).

| Metrics | Region | Rank | | | |
|---|---|---|---|---|---|
| | | **First** | **Second** | **Third** | **Fourth** |
| **Rmse** | Beijing-Tianjin-Hebei | GR-BSReLU | GR-SSReLU | GR-AEReLU | GR-COReLU |
| | Southeastern coastal | GR-BSReLU | GR-AEReLU | GR-SSReLU | GR-COReLU |
| | Sichuan-Yunnan | GR-SSReLU | GR-AEReLU | GR-COReLU | GR-BSReLU |
| | Northern Xinjiang | GR-AEReLU | GR-COReLU | GR-SSReLU | GR-BSReLU |
| | California | GR-BSReLU | GR-AEReLU | GR-SSReLU | GR-COReLU |
| | New Zealand | GR-AEReLU | GR-SSReLU | GR-COReLU | GR-BSReLU |
| **R-Square** | Beijing-Tianjin-Hebei | GR-BSReLU | GR-SSReLU | GR-AEReLU | GR-COReLU |
| | Southeastern coastal | GR-BSReLU | GR-AEReLU | GR-SSReLU | GR-COReLU |
| | Sichuan-Yunnan | GR-SSReLU | GR-AEReLU | GR-COReLU | GR-BSReLU |
| | Northern Xinjiang | GR-AEReLU | GR-COReLU | GR-SSReLU | GR-BSReLU |
| | California | GR-BSReLU | GR-AEReLU | GR-SSReLU | GR-COReLU |
| | New Zealand | GR-AEReLU | GR-SSReLU | GR-COReLU | GR-BSReLU |
| **SSE** | Beijing-Tianjin-Hebei | GR-BSReLU | GR-SSReLU | GR-AEReLU | GR-COReLU |
| | Southeastern coastal | GR-BSReLU | GR-AEReLU | GR-SSReLU | GR-COReLU |
| | Sichuan-Yunnan | GR-SSReLU | GR-AEReLU | GR-COReLU | GR-BSReLU |
| | Northern Xinjiang | GR-AEReLU | GR-BSReLU | GR-SSReLU | GR-COReLU |
| | California | GR-BSReLU | GR-AEReLU | GR-SSReLU | GR-COReLU |
| | New Zealand | GR-AEReLU | GR-SSReLU | GR-COReLU | GR-BSReLU |
| **AIC** | Beijing-Tianjin-Hebei | GR-BSReLU | GR-SSReLU | GR-AEReLU | GR-COReLU |
| | Southeastern coastal | GR-BSReLU | GR-AEReLU | GR-SSReLU | GR-COReLU |
| | Sichuan-Yunnan | GR-SSReLU | GR-AEReLU | GR-COReLU | GR-BSReLU |
| | Northern Xinjiang | GR-AEReLU | GR-BSReLU | GR-SSReLU | GR-COReLU |
| | California | GR-BSReLU | GR-AEReLU | GR-SSReLU | GR-COReLU |
| | New Zealand | GR-AEReLU | GR-SSReLU | GR-COReLU | GR-BSReLU |
| ***p*-value** | Beijing-Tianjin-Hebei | GR-BSReLU | GR-AEReLU | GR-SSReLU | GR-COReLU |
| | Southeastern coastal | GR-SSReLU = GR-COReLU | | GR-AEReLU | GR-BSReLU |
| | Sichuan-Yunnan | GR-AEReLU | GR-COReLU | GR-SSReLU | GR-BSReLU |
| | Northern Xinjiang | GR-SSReLU = GR-BSReLU = GR-COReLU = GR-AEReLU | | | |
| | California | GR-SSReLU = GR-BSReLU = GR-AEReLU | | | GR-COReLU |
| | New Zealand | GR-BSReLU | GR-AEReLU | GR-SSReLU | GR-COReLU |